# Synthesis of the Elusive Bulk Iodide Double-Perovskite Semiconductor, $Cs_2AgBiI_6$: Microcrystals and Photoconductive Films

Faris Horani, Nicolas Nguyen, Carmelita Ro-Mendez, Nicholas J. Adams, and Daniel R. Gamelin*

*Department of Chemistry, University of Washington, Seattle, Washington 98195-1700, United States*

Email: gamelin@uw.edu

**ABSTRACT.** Three-dimensional (3D) iodide double-perovskite (elpasolite) semiconductors have attracted interest as potential lead-free metal-halide absorber layers for solar applications. Although studied extensively by computational methods, they have remained largely inaccessible synthetically, consistent with their predicted thermodynamic instability. Here, we report the first synthesis of bulk $Cs_2AgBiI_6$, demonstrating both microcrystalline and thin-film forms. Microcrystalline powders of $Cs_2AgBiI_6$ were prepared *via* anion exchange from phase-pure $Cs_2AgBiBr_6$ microcrystals. The resulting iodide elpasolite shows broad absorption throughout the visible with a $1.70 \pm 0.05$ eV optical bandgap and near-infrared photoluminescence centered at 1.03 eV. We identify the elimination of trace moisture in bulk $Cs_2AgBiBr_6$ as the critical factor enabling complete halide exchange and isolation of bulk $Cs_2AgBiI_6$ with phase purity. In inert atmosphere, microcrystalline $Cs_2AgBiI_6$ shows no decomposition when stored for months at room temperature or heated to ~70 °C, and it appears equally stable in dry air. Building upon these insights, we then demonstrate the preparation of phase-pure $Cs_2AgBiI_6$ films by flash thermal evaporation of $Cs_2AgBiBr_6$ followed by anion exchange. Photoconductivity measurements on such $Cs_2AgBiI_6$ films demonstrate photocarrier generation and transport, marking the first optoelectronic measurement on this elusive 3D iodide double perovskite.

## Introduction

The development of three-dimensional (3D) iodide double-perovskite (elpasolite) semiconductors represents a longstanding challenge in the quest for lead-free alternatives to lead-halide perovskite semiconductors. Although lead-halide perovskites show exceptional light-harvesting and charge-transport properties,[1] concerns regarding their toxicity and long-term operational stability continue to drive intensive efforts to identify alternative perovskite-like materials suitable for photovoltaics or other optoelectronic applications. 3D double perovskites with the general formula $A_2M^{(I)}M^{(III)}X_6$ (where $M^{(I)}$ is a monovalent cation, $M^{(III)}$ is a trivalent cation, and $X^-$ is a halide) offer broad compositional tunability that has already made them useful for photovoltaics,[2–4] light-emitting devices,[5,6] X-ray detectors,[7] and as hosts for luminescent ions,[8–

[10] but their utility has not yet been fully realized because of the inability to stabilize their iodide form ($A_2M^{(I)}M^{(III)}I_6$). Extensive computational work points to intrinsic thermodynamic instabilities of most iodide elpasolite compositions that might possess intermediate band gaps, as the large ionic radius of $I^-$ requires relatively large $M^{(I)}$ and $M^{(III)}$ cations for lattice stability,[11–15] leaving few suitable $M^{(I)}/M^{(III)}$ cation combinations.

The scope of known 3D iodide elpasolites thus remains small, and most are wide-bandgap insulators. Rare-earth compounds with compositions of $Cs_2NaM^{(III)}I_6$ and $Cs_2LiM^{(III)}I_6$ ($M^{(III)}$ = Sc, Lu, Tm, Er, Ho)[16] were reported in 1980 and assigned to elpasolite structures based on Guinier–Simon powder diffraction. Recently, the elpasolite structures of several $Cs_2NaM^{(III)}I_6$ (M = Ce, Nd, Gd, Tb, Dy) compounds were demonstrated using single-crystal X-ray diffraction and synchrotron powder X-ray diffraction.[17] All of these materials are wide-gap insulators, however. Among 3D iodide elpasolite *semiconductors*, the $A_2Au^{(I)}Au^{(III)}I_6$ compounds (with A = $Cs^+$, $Rb^+$, $K^+$, $CH_3NH_3^+$)[18–27] stand alone as the only compositions with intermediate bandgaps that have been isolated in bulk form, exhibiting "nearly direct" optical gaps of ~0.9 eV (calculated ~1.1–1.3 eV).[28–30]

The paucity of 3D iodide elpasolite semiconductors has motivated interest in the basic chemistry of these materials. Bulk $Cs_2NaBiI_6$ has been reported,[31] but its diffraction data suggest that these samples may instead have been the competing phase, $Cs_3Bi_2I_9$, which is the thermodynamically favored decomposition product. In parallel, our group has reported isolation of $Cs_2AgBiI_6$ and $Cs_2AgSbI_6$ elpasolites,[32,33] both showing intermediate bandgaps of ~1.7-1.8 eV. These bandgaps are well suited for use as top absorber layers pairing with Si in tandem PV configurations. In both cases, however, the iodide phase could only be attained as nanocrystals, formed *via* post-synthetic halide exchange of the corresponding bromide nanocrystals. Despite their compositional and structural similarities, $Cs_2AgBiI_6$ and $Cs_2AgSbI_6$ show markedly different environmental stability: $Cs_2AgBiI_6$ nanocrystals decompose rapidly upon exposure to air, whereas $Cs_2AgSbI_6$ nanocrystals remain stable in air even up to 140 °C, highlighting how cation size can play a decisive role in the phase stability of these iodide elpasolites. Additional studies of $Cs_2AgBiI_6$ nanocrystals have further shown that reduced dimensionality alone is not sufficient to stabilize this lattice, revealing that the crystal surfaces also play a key role, with certain ligands, additives, and ligand loss appearing to favor decomposition under various conditions.[34]

Although isolable as nanocrystals, bulk $Cs_2AgBiI_6$ has to date remained elusive. Efforts to make this material by mechanochemical methods[35,36] found that iodide alloying within $Cs_2AgBi(Br_{1-x}I_x)_6$ was limited to <11 mol %[36] before the accumulation of secondary phases such as $Cs_3Bi_2I_9$. Similar limitations have been encountered in attempts to make bulk $Cs_2AgBiI_6$ by anion exchange. For example, studies of $Cs_2AgBi(Br_{1-x}I_x)_6$ films prepared by reacting spin-coated $Cs_2AgBiBr_6$ films with methylammonium iodide followed by thermal annealing in a nitrogen-filled dry box found that the iodide content could be increased to $x \sim 0.67$ (with a decrease in the optical bandgap from ~2.30 eV to ~2.03 eV), but further iodide exposure led to decomposition into $Cs_3Bi_2I_9$ rather than complete conversion to $Cs_2AgBiI_6$.[37] Similarly, efforts in our own labs to adapt gas-phase halide-exchange strategies that were successful with nanocrystals proved unsuccessful in bulk: Whereas reaction of TMSI with $CsPbBr_3$ films can drive complete halide exchange to form the metastable perovskite-phase "black" $CsPbI_3$,[38] analogous reactions with $Cs_2AgBiBr_6$ films allowed iodide incorporation only up to $Cs_2AgBi(Br_{0.66}I_{0.34})_6$ before decomposition into secondary phases.[34] A study published while this manuscript was under review reports the formation of $Cs_2AgBi(Br,I)_6$ double perovskites *via* solid-state reaction between $Cs_3Bi_2(Br,I)_9$ and $CsAg_2I_3$ at 200–300 °C, with maximum iodide levels ~88%.[39] Under optimized conditions, the products of this reaction consisted predominantly of tetragonal $Cs_2AgBi(Br,I)_6$, accompanied by residual $Cs_3Bi_2(Br,I)_9$ and $CsAg_2I_3$. Bulk $Cs_2AgBiI_6$ was not formed under these conditions, and the stability of the mixed-halide $Cs_2AgBi(Br,I)_6$ compositions was attributed to the presence of bromide in the lattice. Collectively, these unsuccessful outcomes have all reinforced the view that bulk $Cs_2AgBiI_6$ is inaccessible.

In this work, we report the first preparation of bulk $Cs_2AgBiI_6$. Two forms of this material have been studied: microcrystalline powders and thermally evaporated films. Using phase-pure $Cs_2AgBiBr_6$ microcrystals prepared by mechanochemical synthesis, we show that reaction with TMSI vapor under inert atmosphere yields complete anion exchange without decomposition or formation of impurity phases. The resulting $Cs_2AgBiI_6$ microcrystals show broad-band absorption with a $1.70 \pm 0.05$ eV bandgap, and near-infrared photoluminescence at 1.03 eV. This $Cs_2AgBiI_6$ is unstable in humid air but is stable under nitrogen or dry air for at least months with no sign of decomposition. Critically, we find that the synthesis of bulk $Cs_2AgBiI_6$ is highly sensitive to trace water: Although $Cs_2AgBiBr_6$ is often described as air stable, $Cs_2AgBiBr_6$ powders exposed to air adsorb moisture that even in small quantities impedes anion exchange by catalyzing decomposition into secondary phases at low iodide content. This observation likely explains the challenges faced in previous attempts to prepare bulk $Cs_2AgBiI_6$. We show that annealing air-exposed $Cs_2AgBiBr_6$ anaerobically before anion exchange dehydrates the surfaces and allows complete conversion to impurity-free $Cs_2AgBiI_6$. Armed with this insight, we then prepared phase-pure $Cs_2AgBiI_6$ films *via* thermal evaporation of $Cs_2AgBiBr_6$ followed by gas-phase halide exchange. Preparation of such films directly on interdigitated electrodes has allowed demonstration of promising broadband photoconductivity throughout the visible. Together, these results establish a robust and scalable route to bulk and thin-film $Cs_2AgBiI_6$, resolving critical synthetic challenges and highlighting the attractive properties of this extremely rare example of a 3D iodide elpasolite semiconductor.

**Results and Discussion**

Although experimental data for 3D iodide elpasolites are rare, geometric descriptors still offer an informative starting point for evaluating their structural viability. A common metric for assessing the stability of perovskites is the Goldschmidt tolerance factor (**eq 1**).[40]

$$t = \frac{R_A + R_X}{\sqrt{2}(\langle R_M \rangle + R_X)} \tag{1}$$

Here, $R_A$, $R_X$, and $R_M$ denote A-site, halide, and metal ionic radii.[41] In eq 1, the effective $R_M$ for a double perovskite is taken as the arithmetic mean of the two B-site cation radii ($\langle R_M \rangle = \frac{1}{2}(R_{M(I)} + R_{M(III)})$). For $Cs_2AgBiI_6$, eq 1 yields $t = 0.88$, placing this material within the canonical stability window for halide perovskites ($0.80 \le t \le 1.00$). Its octahedral factor, $\langle R_M \rangle / R_X = 0.50$, is also within the acceptable window (from 0.41 to 0.90). Although broadly useful, however, this hard-sphere geometric model becomes less predictive for lattices containing highly polarizable anions such as $I^-$.[42] To provide a more discriminating evaluation, we may also employ the recently revised tolerance factor $t'$ (**eq 2**),[13] which incorporates the A-site oxidation state ($n_A$) and captures electrostatic and bond-valence effects that are absent from the purely geometric Goldschmidt description. Within this framework, $Cs_2AgBiI_6$ ($t'$ = 4.18) resides at the very edge of the predicted stability domain for perovskites ($t'$ < 4.18). Taken together, these metrics suggest that bulk $Cs_2AgBiI_6$ may indeed occupy a narrow yet accessible stability window, providing the impetus for the synthetic and structural investigations described here.

$$t' = \frac{R_X}{\langle R_M \rangle} - n_A\left(n_A - \frac{\frac{R_A}{\langle R_M \rangle}}{\ln\left(\frac{R_A}{\langle R_M \rangle}\right)}\right) \tag{2}$$

We first explored the preparation of bulk $Cs_2AgBiBr_6$ using microcrystalline powders (see Experimental section for details). Briefly, microcrystalline $Cs_2AgBiBr_6$ was synthesized by wet ball milling of CsBr, AgBr, and $BiBr_3$ in toluene for 8 h at 550 rpm, followed by filtration and washing with toluene. Annealing this product at 250 °C under inert atmosphere yielded highly crystalline, phase-pure $Cs_2AgBiBr_6$ **(Figure S1)**. To convert this product to $Cs_2AgBiI_6$, 75 mg of the $Cs_2AgBiBr_6$ microcrystals was sealed for 24 h in a 20 mL vessel alongside a separate vial containing 500 μL of TMSI. During this time, the powders changed from yellow to black, indicating anion exchange.

For film preparation, annealed $Cs_2AgBiBr_6$ powders were used as a single-source precursor for thermal evaporation onto silicon or other substrates, followed by a 30-min anneal at 250 °C under inert atmosphere to promote grain growth, remove residual precursors, and eliminate adventitious moisture. These films were then reacted with TMSI vapor for 24 h under the same conditions as above, resulting in a color change from yellow to dark red. **Figure 1** summarizes the full procedures for preparing microcrystalline and thin-film $Cs_2AgBiI_6$. The resulting materials were stored under inert atmosphere for subsequent characterization.

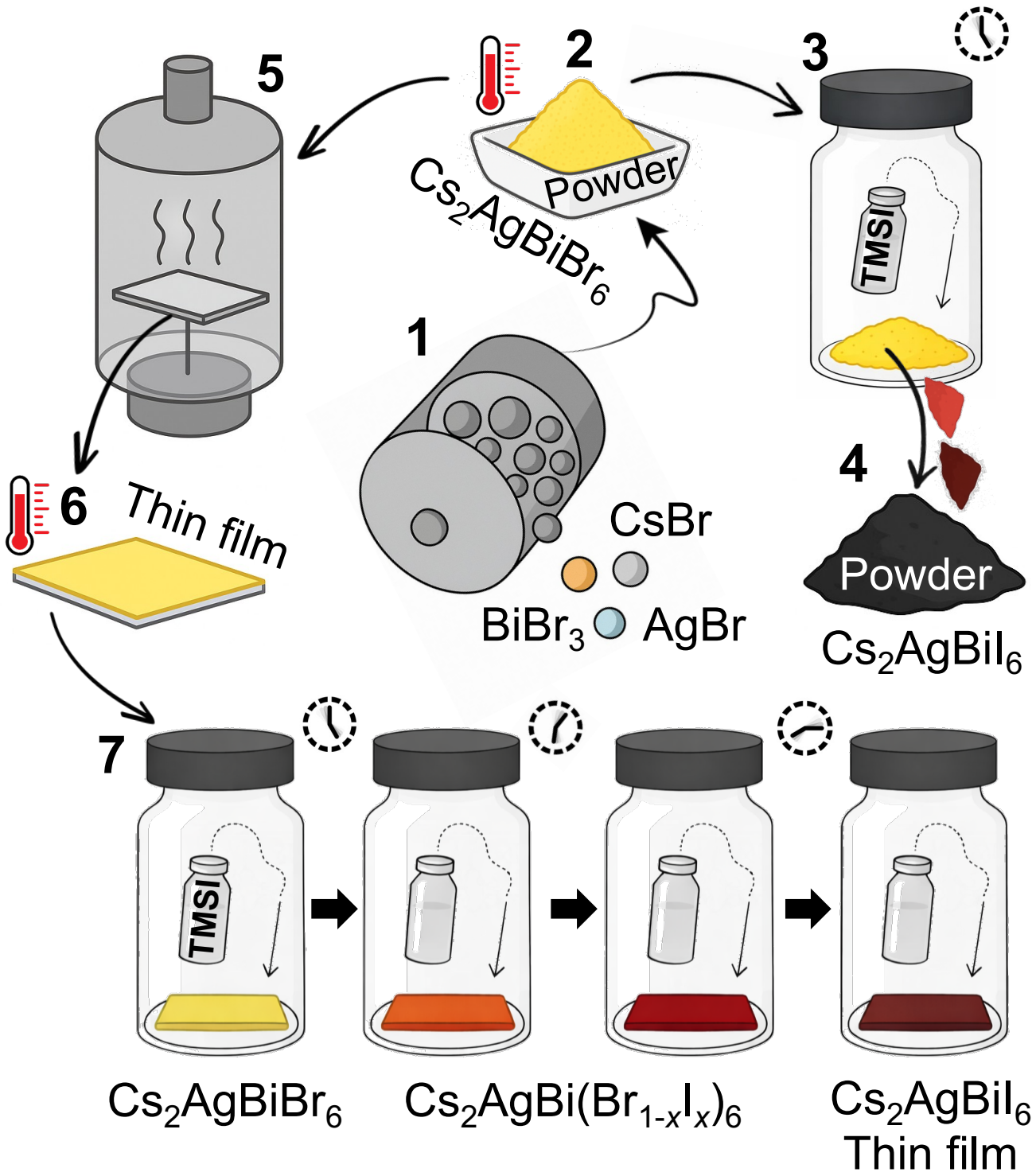

**Figure 1.** Schematic overview of the synthesis routes for $Cs_2AgBiI_6$ powders and films. **Powders:** $Cs_2AgBiBr_6$ is first prepared by ball milling (Step 1), annealed at 250 °C (Step 2), and subsequently converted to $Cs_2AgBiI_6$ *via* vapor-phase reaction with TMSI (Step 3),

during which the material changes color from yellow to black. **Films:** The annealed powder (Steps 1–2) is used as the precursor for thermal evaporation to form $Cs_2AgBiBr_6$ films (Step 5), followed by annealing under inert atmosphere (Step 6) and reaction with TMSI vapor (Step 7), resulting in a color change from yellow to dark red.

**Figure 2A, B** shows representative photographs of microcrystalline and thin-film elpasolite samples at various stages of the anion-exchange process. Video recordings (Supporting Information, **Video S1**) show rapid initial color changes in both the powder and film samples that then appear by eye to approach completion after ~3 h of TMSI exposure. To ensure complete halide exchange throughout the interior volumes of the microcrystals and film grains, samples were kept in TMSI vapor for 24 h. After 24 h, the samples were placed under dynamic vacuum to remove any remaining TMSBr or TMSI. Note that the driving force for $TMSI_{gas} + Br^{-}_{surf} \rightarrow TMSBr_{gas} + I^{-}_{surf}$ is large, such that this reaction is (gas-phase) diffusion limited and effectively irreversible.[32,34,38] The rate of the solid's overall anion-exchange reaction is therefore governed by anion diffusion within the crystal lattice itself. For this reason, the overall conversion is markedly slower in these microcrystalline samples than in the elpasolite nanocrystals studied previously.[32,34]

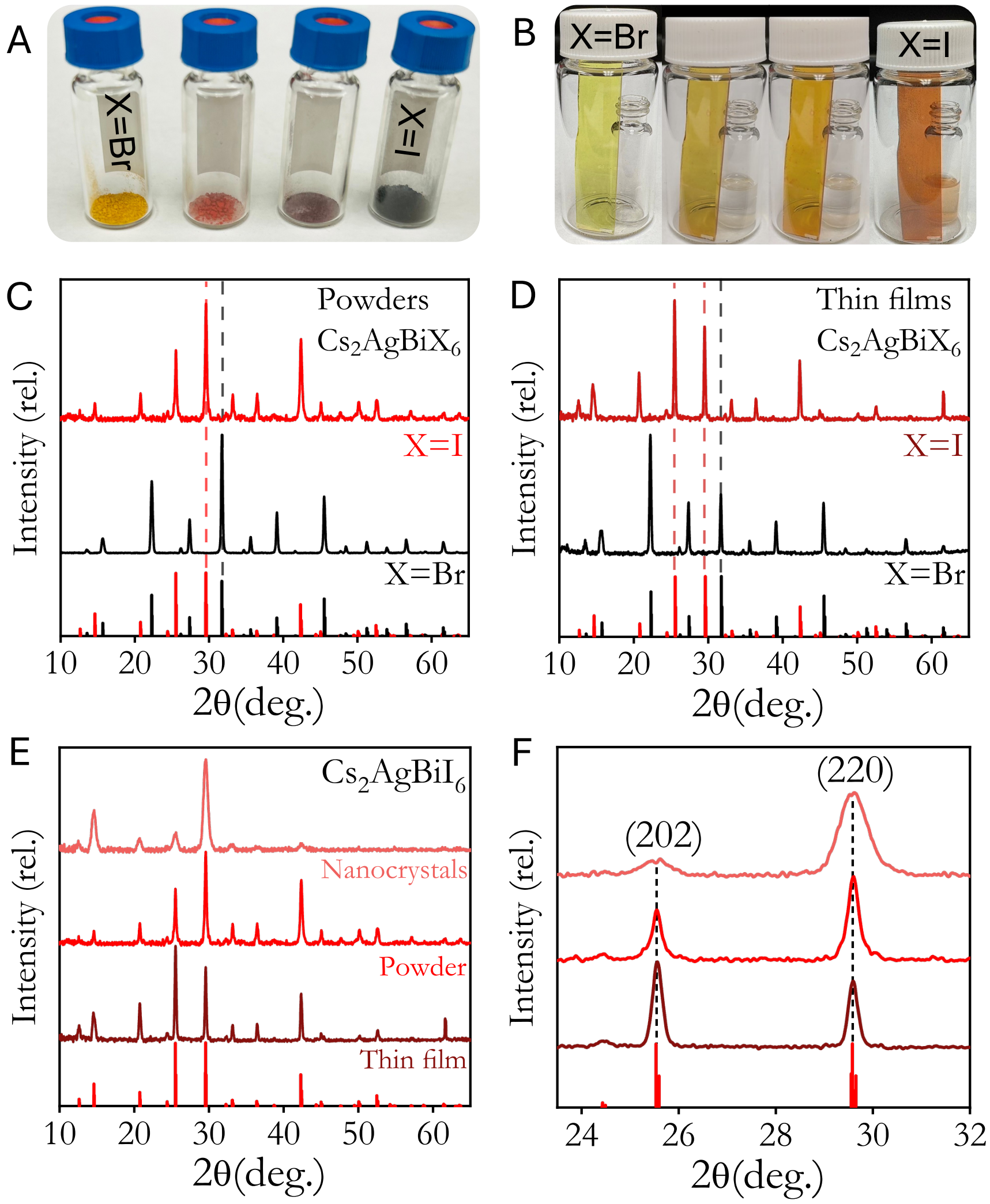

**Figure 2. (A)** Photographs illustrating the conversion of yellow $Cs_2AgBiBr_6$ powder (left) into black $Cs_2AgBiI_6$ powder (right) upon exposure to TMSI vapor. Exposure times = 0, 30 min, 1 h, and 24 h. **(B)** Photographs of films during the same anion-exchange procedure (left to right). TMSI exposure times = 0, 30 min, 2 h, and 24 h. Note that the apparent coloration of the TMSI vial is only an optical effect from the presence of the colored film. **(C, D)** XRD patterns of $Cs_2AgBiBr_6$ and $Cs_2AgBiI_6$ powders and films. Reference powder XRD patterns for $Cs_2AgBiBr_6$ and $Cs_2AgBiI_6$ are shown for comparison. **(E)** Overlaid XRD patterns of $Cs_2AgBiI_6$ films, powders, and nanocrystals, with grain sizes decreasing from largest to smallest, respectively. Reference powder XRD patterns for $Cs_2AgBiBr_6$ and $Cs_2AgBiI_6$ are shown for comparison. **(F)** Magnified XRD region emphasizing the absence of secondary phases and the modest preferential (202) relative to (220) orientation in the $Cs_2AgBiI_6$ film. Reference powder XRD pattern for $Cs_2AgBiI_6$ is shown for comparison.

Powder XRD data for the $Cs_2AgBiBr_6$ microcrystals and films exposed to TMSI both show shifts of all major reflections to lower 2θ angles, consistent with lattice expansion due to substitution of smaller $Br^-$ by larger $I^-$ anions (**Figures 2C, D**). No secondary or exsolved phases were detected, even when plotting the intensities on a logarithmic scale (**Figure S2**). In particular, the characteristic reflection of the decomposition product $Cs_3Bi_2I_9$ at ~27.5° is absent; $Cs_2AgBiI_6$ and $Cs_3Bi_2I_9$ XRD patterns are very similar due to their related unit cells, but this unique peak at 27.5° provides a sensitive and selective measure of the presence of $Cs_3Bi_2I_9$.[34] To estimate the $Cs_3Bi_2I_9$ content quantitatively, reference samples were made by mixing known masses of $Cs_2AgBiI_6$ and $Cs_3Bi_2I_9$ powders, from which a calibration plot of relative $Cs_3Bi_2I_9$ (203) to [$Cs_3Bi_2I_9$ (203) + $Cs_2AgBiI_6$ (220)] XRD intensities versus $Cs_3Bi_2I_9$ content was generated (**Figure S3**). This calibration plot and the baseline noise of the $Cs_2AgBiI_6$ XRD data yield a $Cs_3Bi_2I_9$ impurity level of 0.0 ± 0.5 wt%, from which we conclude phase purity. $Cs_3Bi_2I_9$ is the universal decomposition product of $Cs_2AgBiI_6$, forming stoichiometrically with either $CsAg_2I_3$ or 2AgI + CsI depending on conditions. None of these other potential impurities are detected, either.

The diffraction profiles of the $Cs_2AgBiI_6$ powders and films both closely match the standard diffraction of the tetragonal (*I*4/*m*) elpasolite phase (**Figure 2E, F**) determined previously from Rietveld refinement of synchrotron XRD data,[34] as well as the experimental diffraction data from 13 nm $Cs_2AgBiI_6$ nanocrystals prepared independently. In the films, the intensity of the 25.5° reflection, attributed to the (202) plane, is enhanced relative to the 29.5° reflection (220), suggesting a slightly preferred orientation. The diffraction peaks in the microcrystal and thin-film samples are substantially narrower than those of the nanocrystals, reflecting much greater structural correlation lengths. We thus conclude the successful preparation of bulk $Cs_2AgBiI_6$.

Rietveld refinement was performed using the powder XRD data from a representative bulk microcrystalline $Cs_2AgBiI_6$ sample, and the results are summarized in **Figure S4** of the Supporting Information. An initial structural model of tetragonal $Cs_2AgBiI_6$ (*I*4/*m*) was adapted from nanocrystalline $Cs_2AgBiI_6$.[34] The refinement converged to a high-quality fit with $R_w$ = 5.75% and a goodness-of-fit of $\chi^2$ = 1.1. These results show that bulk $Cs_2AgBiI_6$ has the same tetragonal structure as nanocrystalline $Cs_2AgBiI_6$.

Scanning electron microscopy (SEM) images (**Figure 3A, Figure S5**) of the $Cs_2AgBiI_6$ powders show faceted crystallites of many sizes, with an average length approaching 1 μm. Such broad size distributions are common for mechanochemically synthesized bulk powders.[43] These images closely resemble those of the $Cs_2AgBiBr_6$ precursor powders (**Figure S6**), consistent with the preservation of crystal morphology during anion exchange that was observed in the analogous nanocrystal reactions.[32,34] Energy-dispersive X-ray spectroscopy (EDX) of the $Cs_2AgBiI_6$ powders shows no detectable signals from Br ($L_\alpha \approx 1.48$ keV, $K_\alpha \approx 11.92$ keV). The elemental mapping is uniform across the entire probe area, and cation ratios are consistent with the elpasolite composition (**Figure S5, Table S1**). X-ray photoelectron spectroscopy (XPS) corroborates these findings, revealing characteristic orbital splittings corresponding to $Cs^+$, $Ag^+$, $Bi^{3+}$, and $I^-$ (**Figure 3C**). From these results, we conclude complete halide exchange to form bulk $Cs_2AgBiI_6$ with no detectable residual $Br^-$. Similar results are found for films: SEM images of representative $Cs_2AgBiI_6$ films show dense material over large areas, with individual grain sizes on the order of ~1 μm. An average film thickness of ~200 nm was estimated by SEM (**Figure 3B, Figures S7-9, Table S1**).

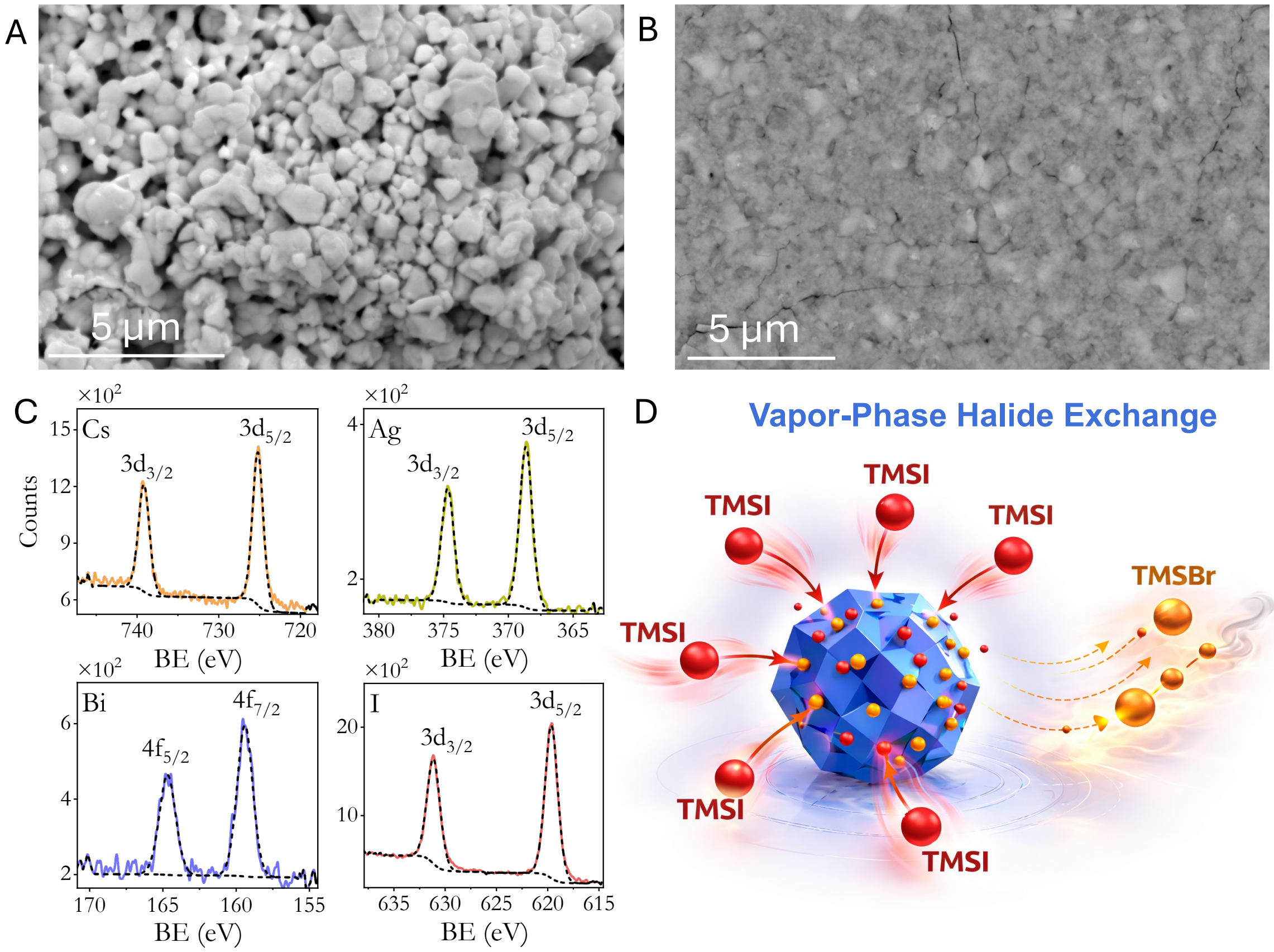

**Figure 3.** SEM images of **(A)** $Cs_2AgBiI_6$ powder and **(B)** a representative $Cs_2AgBiI_6$ film. **(C)** XPS spectra highlighting the Cs, Ag, Bi, and I core-level 3*d* and 4*f* transitions of $Cs_2AgBiI_6$ powder. **(D)** Schematic illustration of the topotactic anion-exchange conversion of $Cs_2AgBiBr_6$ into $Cs_2AgBiI_6$ upon exposure to TMSI.

Diffuse reflectance spectroscopy (DRS) was used to compare the optical properties of $Cs_2AgBiI_6$ with those of $Cs_2AgBiBr_6$ and $Cs_3Bi_2I_9$. **Figure 4A** plots the Kubelka–Munk function, F(*R*), measured for powders of these three materials at room temperature. The $Cs_2AgBiI_6$ spectrum shows the narrowest optical gap, with continuous broadband absorption throughout the visible spectral range. Analysis of Tauc plots (**Figure 4B**) indicates indirect bandgap transitions for $Cs_2AgBiI_6$, $Cs_2AgBiBr_6$, and $Cs_3Bi_2I_9$, with optical gaps of 1.70, 2.30, and 2.10 eV (~730, 540, and 590 nm), respectively, and an estimated fitting uncertainty of ±0.05 eV. $Cs_2AgBiI_6$ also shows an Urbach-like absorption tail that extends down to ~1.4 eV.

The $Cs_2AgBiI_6$ powder shows Stokes-shifted near-infrared photoluminescence (PL) at 4 K, centered at 1.03 eV (~1200 nm) with a full-width-at-half-maximum (FWHM) of ~0.30 eV (**Figure 4C**). This PL redshifts slightly with increasing temperature. Its intensity grows as the temperature is raised from 4 to ~50 K, consistent with thermal detrapping from shallow traps, before decreasing again at higher temperatures because of nonradiative recombination (**Figure 4D**). Weak PL is still observed at room temperature (**Figure S10**). Time-resolved PL measurements show multiexponential decay that is described well by a tri-exponential model (**Figure S11** and **Table S2**), with an average decay time of $\tau \approx$ 10.7 μs at 4 K. $\tau$ increases slightly with increasing temperature, peaking at ~13 μs near 80 K before decreasing again at higher temperatures and dropping to ~440 ns at 270 K (**Figures 4E, F**). Like the intensity data, these PL decay data are also consistent with thermal detrapping in the low-temperature range and enhanced nonradiative recombination at elevated temperatures. These data suggest assignment of this PL as coming from deep trapping at defects, although self-trapping cannot be ruled out.

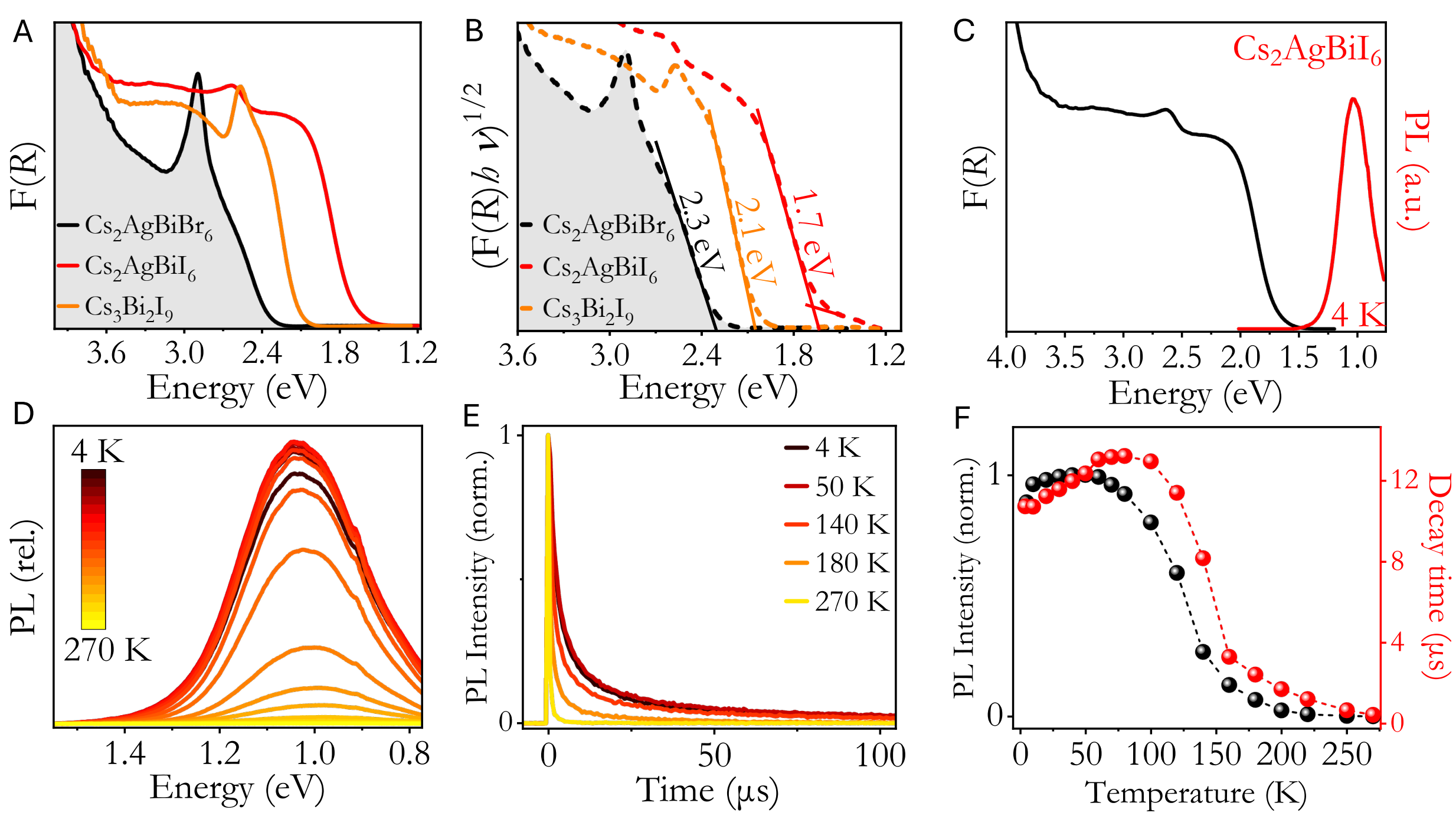

**Figure 4. (A)** Room-temperature diffuse reflectance spectra of bulk $Cs_2AgBiBr_6$, $Cs_3Bi_2I_9$, and $Cs_2AgBiI_6$ powder samples, presented as the Kubelka–Munk function F(*R*) *vs* energy. **(B)** Tauc plots of $(F(R)h\nu)^2$ *vs* energy for the samples from (A), showing the linear fits used to estimate the optical bandgaps. The fitting uncertainties are estimated to be ±0.05 eV. **(C)** Diffuse reflectance spectrum and 4 K PL spectrum of bulk $Cs_2AgBiI_6$. **(D)** Variable-temperature PL spectra of bulk $Cs_2AgBiI_6$ recorded from 4 to 270 K under 405 nm excitation. **(E)** Time-resolved PL traces of the emission centered at 1.03 eV as function of temperature, excited using 355 nm laser pulses. **(F)** Temperature dependence of the integrated PL intensity (black) and average PL decay time (red) from panels D and E.

The isolation of bulk $Cs_2AgBiI_6$ stands in stark contrast with several prior observations of secondary-phase formation at intermediate iodide concentrations,[34–37] including from our own group using similar anion-exchange chemistry.[34] To understand this result, we scrutinized the synthesis, handling, and anion-exchange reaction conditions and discovered that exposure of $Cs_2AgBiBr_6$ to air is the single most critical factor in our procedure that compromises the ability to isolate bulk $Cs_2AgBiI_6$ without decomposition.

To illustrate, Figure 5 compares the results from two parallel sample-processing paths that differ only in the atmosphere under which the ball-milled $Cs_2AgBiBr_6$ is annealed (step 2 of Figure 1). **Figure 5A, B** plots X-ray diffraction data showing that $Cs_2AgBiBr_6$ annealed in nitrogen and then reacted with TMSI under inert atmosphere yields phase-pure $Cs_2AgBiI_6$, whereas $Cs_2AgBiBr_6$ annealed in ambient air and then reacted with TMSI under inert atmosphere and identical conditions forms $Cs_3Bi_2I_9$ and $CsAg_2I_3$, *even though both $Cs_2AgBiBr_6$ starting materials are phase pure*. Raman spectra of $Cs_2AgBiI_6$ formed by reaction of TMSI with $Cs_2AgBiBr_6$ annealed in nitrogen (**Figure 5C**) show vibrations at 130, 105, and 59 $cm^{-1}$, assigned to the $A_{1g}$, $E_g$, and $T_{2g}$ modes of $[BiI_6]^{3-}$ octahedra.[44] These features closely resemble those of $Cs_2AgBiBr_6$ but are all red-shifted, consistent with iodide substitution. In contrast, the product obtained from reacting air-annealed $Cs_2AgBiBr_6$ with TMSI shows only the characteristic Raman features of $Cs_3Bi_2I_9$ (**Figure 5D**), including bands at 145 and 118 $cm^{-1}$ associated with terminal Bi–I stretching, peaks at 104 and 89 $cm^{-1}$ corresponding to bridging Bi–I vibrations, and a low-frequency mode at 56 $cm^{-1}$ assigned to Bi–I bending.[44]

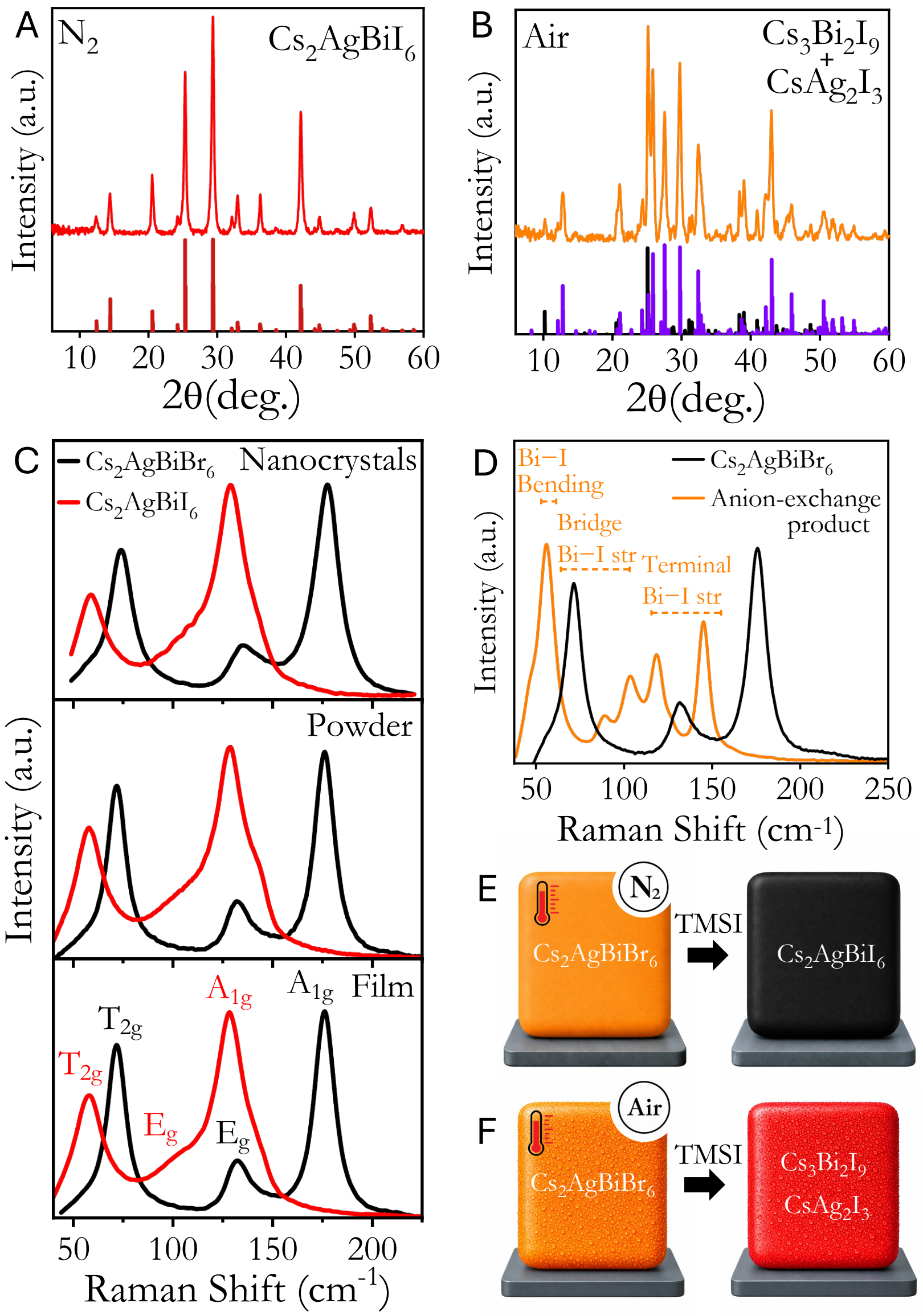

**Figure 5. (A, B)** XRD patterns of $Cs_2AgBiBr_6$ powders annealed in nitrogen and air following TMSI-mediated anion exchange. **(C)** Raman spectra of $Cs_2AgBiBr_6$ and $Cs_2AgBiI_6$ nanocrystals (top), microcrystalline powder (middle), and films (bottom), showing the characteristic vibrational signatures of the $Cs_2AgBiI_6$ elpasolite phase following anion exchange under $N_2$ atmosphere. For powders and films, the $Cs_2AgBiBr_6$ precursors were annealed under $N_2$ prior to TMSI exposure. **(D)** Raman spectra of $Cs_2AgBiBr_6$ annealed in air and the product obtained after reaction of this sample with TMSI. The latter spectrum is dominated by Raman features from $Cs_3Bi_2I_9$, indicating sample decomposition. **(E, F)** Schematic illustration showing complete anion exchange to form $Cs_2AgBiI_6$ after annealing and reacting $Cs_2AgBiBr_6$ under nitrogen, contrasting the decomposition with anion exchange observed after annealing $Cs_2AgBiBr_6$ in ambient air prior to reaction under inert atmosphere. The latter result is attributed to surface hydration of $Cs_2AgBiBr_6$ in air.

We attribute the different reactivities shown in Figure 5 to the effects of surface hydration. Although $Cs_2AgBiBr_6$ is often described as air stable,[2,45,46] bromide elpasolites are generally hygroscopic and readily adsorb water from ambient air.[47–49] Indeed, recent work[50] has demonstrated structural degradation of $Cs_2AgBiBr_6$ over long air-exposure times, as probed by field-emission SEM. In our samples, SEM imaging reveals that $Cs_2AgBiBr_6$ powders annealed under nitrogen show nearly isotropic morphologies and homogeneous elemental distributions, whereas the same powders annealed in air develop microrod-like morphologies (**Figure S12**). Although powder XRD still shows phase-pure $Cs_2AgBiBr_6$ after annealing in air, EDX elemental mapping appears to show spots with higher Ag concentration (**Figure S12**), and these air-annealed samples slowly decompose to $Cs_3Bi_2Br_9$ and AgBr even when stored under inert atmosphere (**Figure S13**).

Importantly, failure of the anion-exchange chemistry is not limited to precursors that have been annealed in air. Careful inspection shows that even brief exposure of $N_2$-annealed bulk $Cs_2AgBiBr_6$ to air at room temperature leads to moisture uptake that is detectable by FTIR (**Figure S14**). This adsorbed water is sufficient to prohibit formation of phase-pure $Cs_2AgBiI_6$ by reaction with TMSI, even when the anion exchange is performed under inert atmosphere (**Figure S15**). Notably, we find that re-annealing hydrated $Cs_2AgBiBr_6$ under inert atmosphere, even if already partially decomposed, restores phase purity as well as the material's capacity to undergo complete iodide exchange to form $Cs_2AgBiI_6$ (**Figures S16, 17**), indicating that the hydration of $Cs_2AgBiBr_6$ is reversible under these conditions.

The above observations, summarized schematically in **Figure 5E, F**, point to surface hydration as the critical factor hindering successful formation of bulk $Cs_2AgBiI_6$ *via* anion exchange using TMSI. The same conclusion likely applies to other methods that have been explored for making bulk $Cs_2AgBiI_6$. Accordingly, strictly anhydrous processing conditions are essential for obtaining phase-pure $Cs_2AgBiI_6$ in its bulk form.

Once prepared, bulk $Cs_2AgBiI_6$ shows remarkable stability under inert conditions. **Figure 6A** compares color photographs of freshly prepared $Cs_2AgBiI_6$, $Cs_2AgBiI_6$ stored under inert atmosphere for months, and $Cs_2AgBiI_6$ exposed to air for one day. XRD data collected after five months of inert storage show no detectable change (**Figure S18**). Upon exposure to ambient air at room temperature, the $Cs_2AgBiI_6$ powders gradually turn from black to red accompanied by a blue shift of the absorption onset (**Figures S19-21**), reflecting decomposition. Importantly, $Cs_2AgBiI_6$ powders stored under ambient atmosphere in the presence of a calcium sulfate desiccant also show no signs of decomposition after several weeks, whereas samples exposed in the same way to ambient air without desiccation decompose within hours (**Figure S22**). This contrast demonstrates that dry $O_2$ does not noticeably degrade $Cs_2AgBiI_6$.

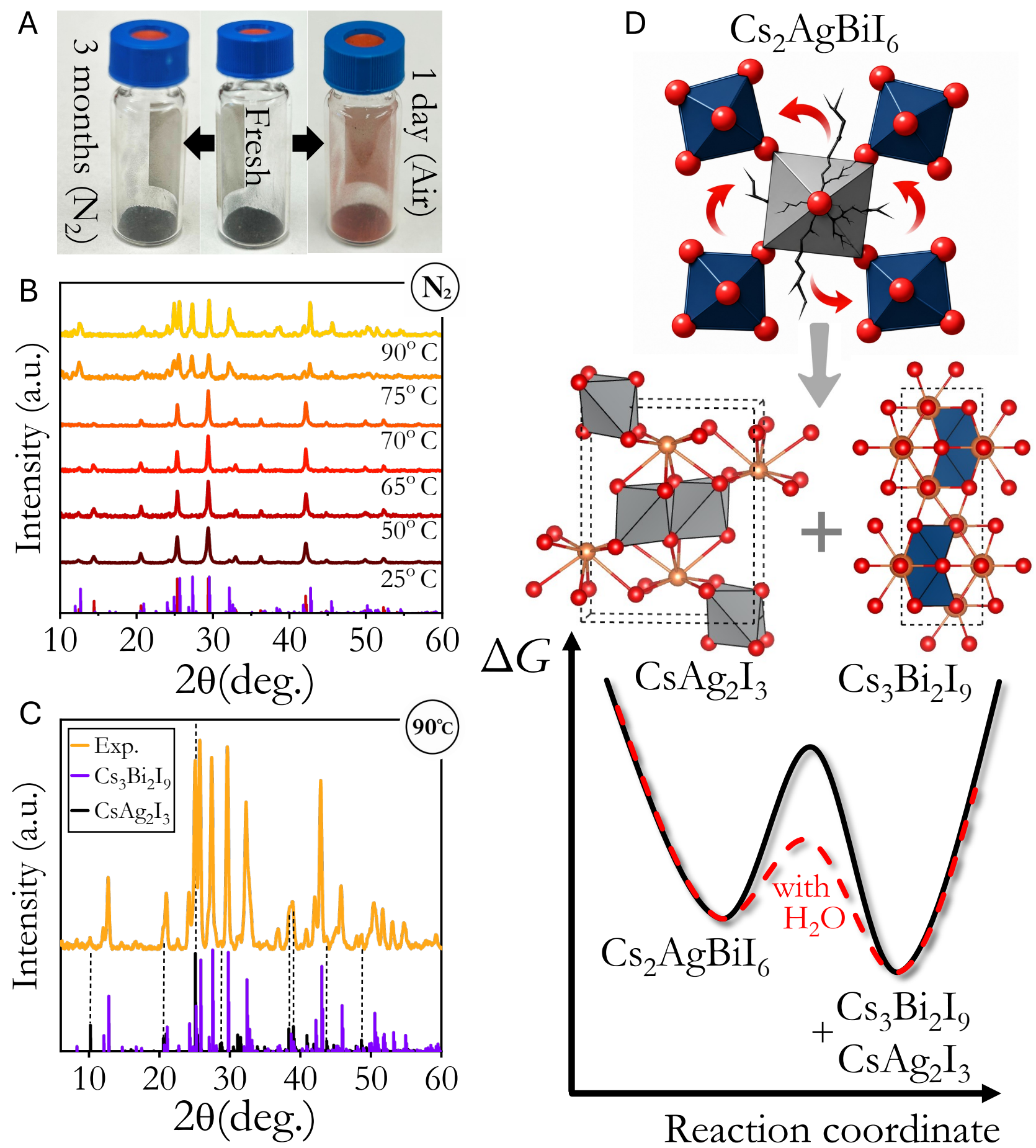

**Figure 6. (A)** Photographs of $Cs_2AgBiI_6$ powders: freshly prepared (middle), stored in inert atmosphere for 3 months (left), and exposed to air for 1 day (right). **(B)** XRD data for $Cs_2AgBiI_6$ powder heated under $N_2$ stepwise (1 h at each temperature) to 90 °C, showing the onset of phase decomposition at ~75 °C. The reference patterns at the bottom correspond to $Cs_2AgBiI_6$ (red) and $Cs_3Bi_2I_9$ (purple). **(C)** The XRD data for $Cs_2AgBiI_6$ powder heated at 90 °C for 1 h from panel B, showing features consistent with $Cs_3Bi_2I_9$ (ICSD 23124) and $CsAg_2I_3$ (ICSD 150308). **(D)** (top) Crystal structures of tetragonal $Cs_2AgBiI_6$, orthorhombic $CsAg_2I_3$, and hexagonal $Cs_3Bi_2I_9$. The gray arrow represents the observed decomposition pathway for $Cs_2AgBiI_6$. (bottom) Gibbs free-energy diagram showing $Cs_2AgBiI_6$ as a metastable but kinetically trapped phase (black curve). $Cs_2AgBiI_6$ decomposes to the thermodynamic products, $Cs_3Bi_2I_9$ and $CsAg_2I_3$, upon heating to ~75

°C. Water adsorbed from air catalyzes this decomposition at room temperature by increasing cation mobility, represented schematically by the lower barrier to decomposition in the red dashed curve.

The thermal stability of $Cs_2AgBiI_6$ was also evaluated by heating powder samples under $N_2$ for one-hour periods at incrementally increasing temperatures. **Figure 6B** shows XRD data revealing that this elpasolite remains stable up to ~70 °C, with decomposition first detected near 75 °C and only clearly evident around 90 °C. XRD analysis of the degraded sample shows $Cs_3Bi_2I_9$ and $CsAg_2I_3$ as the dominant decomposition products (**Figure 6C**). First-principles calculations have proposed two thermodynamically favorable decomposition pathways: 2 $Cs_2AgBiI_6 \rightarrow Cs_3Bi_2I_9$ + 2 AgI + CsI ($\Delta H_r = -0.41$ eV/f.u., calculated with PBEsol)[51] and 2 $Cs_2AgBiI_6 \rightarrow Cs_3Bi_2I_9$ + $CsAg_2I_3$ ($\Delta H_r = -0.48$ eV/f.u., calculated with PBE).[12] The above experimental observations are consistent with dominance of the latter pathway under these conditions.

Despite the first-principles predictions of negative decomposition enthalpies and the past difficulties in achieving stable bulk $Cs_2AgBiI_6$, the results presented above demonstrate that this material can indeed be isolated and is remarkably stable. This apparent discrepancy between theory and experiment may reflect the scope of the first-principles calculations, which capture only the enthalpic driving force for decomposition of the infinite single crystal under idealized conditions but do not account for kinetics, entropy, surfaces and defects, or various other experimentally relevant factors. Nevertheless, our data are consistent with $Cs_2AgBiI_6$ being a metastable "kinetic" product of the anion-exchange reaction between $Cs_2AgBiBr_6$ and TMSI. We hypothesize that the robust kinetic stability of this metastable structure can be attributed to the massive cation rearrangement required to transform $Cs_2AgBiI_6$ into its decomposition products, $Cs_3Bi_2I_9$ and $CsAg_2I_3$. $Cs_3Bi_2I_9$ has a zero-dimensional hexagonal structure composed of discrete $[Bi_2I_9]^{3-}$ dimers, and $CsAg_2I_3$ has a one-dimensional chain-like structure formed by corner-sharing $AgI_4$ tetrahedra, as illustrated in **Figure 6D**. The large kinetic barrier to decomposition is illustrated schematically in the reaction coordinate diagram of **Figure 6D**. We further propose that adsorbed water accelerates decomposition by increasing cation mobility, and in this sense it catalyzes the decomposition, represented by the lowered barrier height to decomposition shown in **Figure 6D**.

The picture that emerges, combining this source of metastability with the ability of water to catalyze decomposition and the ubiquity of trace water, helps us understand the prior difficulties with isolation of bulk $Cs_2AgBiI_6$ by either direct or anion-exchange routes.[34–37] This interpretation of the data even sheds light on the sensitivity of $Cs_2AgBiBr_6$ nanocrystal anion-exchange reactions to different ligands and additives that commonly contain ppm water impurities.[34] This behavior additionally invites comparison with "black"-phase γ-$CsPbI_3$, another metastable iodide-perovskite semiconductor that is kinetically trapped and highly sensitive to moisture.[52,53] For γ-$CsPbI_3$, exposure to humid atmosphere causes rapid conversion to the thermodynamically stable non-perovskite δ-$CsPbI_3$ phase. Despite the similar environmental sensitivity of γ-$CsPbI_3$ and $Cs_2AgBiI_6$, the decomposition pathways of these two materials are fundamentally different. In $CsPbI_3$, the γ-to-δ phase transition occurs without any change in stoichiometry, requiring only a polymorphic rearrangement of the $PbI_6$ framework. Under rigorously anhydrous conditions, γ-$CsPbI_3$ can be kinetically trapped for at least a month at room temperature.[52,53] In contrast, decomposition of $Cs_2AgBiI_6$ involves true chemical breakdown of the lattice into $Cs_3Bi_2I_9$ and $CsAg_2I_3$. The massive cation rearrangement required for this decomposition appears to make $Cs_2AgBiI_6$ more deeply kinetically trapped than γ-$CsPbI_3$. We note that various strategies including doping and surface modification have been successful in improving the kinetic stability of γ-$CsPbI_3$,[54–56] and we anticipate that similar strategies will also allow further stabilization of $Cs_2AgBiI_6$.

With a methodology in hand for preparing this elusive iodide elpasolite in bulk form, it is of interest to explore the extension of this chemistry to a form that could be relevant to optoelectronics. To this end, we have developed a two-step process for fabricating $Cs_2AgBiI_6$ films involving $Cs_2AgBiBr_6$ thermal evaporation followed by anion exchange. After initial proof-of-concept experiments on soda-lime glass substrates, we turned to film deposition onto fused silica substrates patterned with interdigitated gold (20 μm finger spacing) or glassy carbon (100 μm spacing) electrodes to form planar photoconductive devices (see the Experimental section in the Supporting Information for details). **Figure 7A-D** shows photographs of devices based on ~200 nm thick $Cs_2AgBiBr_6$ and $Cs_2AgBiI_6$ films. The $Cs_2AgBiI_6$ film appears brown, rather than the black color of the bulk powder, but its absorption spectrum is the same as that of $Cs_2AgBiI_6$ nanocrystals (**Figure S23**), showing that this color difference comes from the film's thinness. Thicker films could be prepared by adjusting the single-source precursor mass or evaporating a second time onto the same sample.

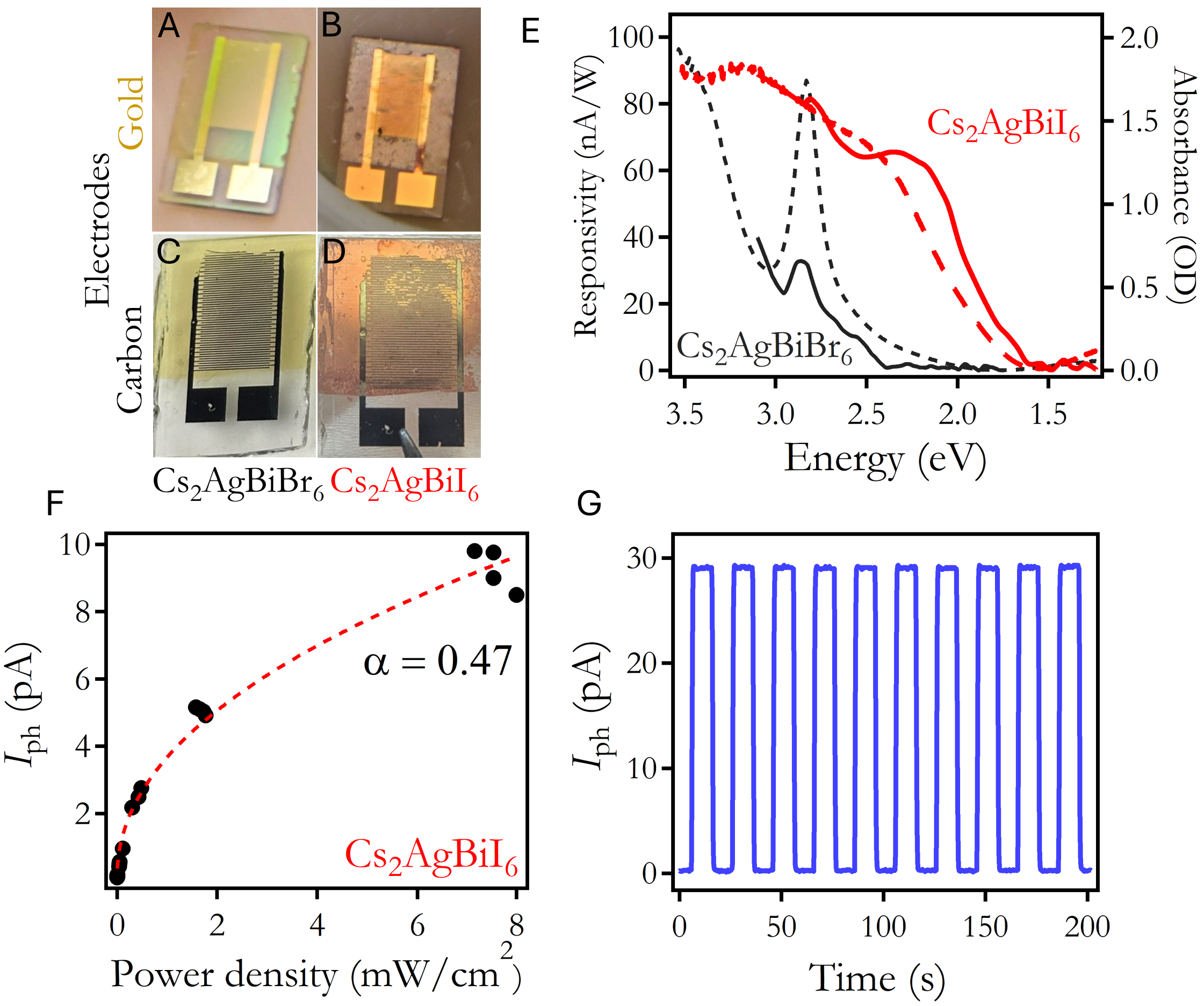

**Figure 7. (A, B)** Photographs of $Cs_2AgBiBr_6$ and $Cs_2AgBiI_6$ films deposited onto fused silica substrates patterned with interdigitated gold electrodes (20 μm spacings). **(C, D)** Photographs of an analogous $Cs_2AgBiBr_6$ film deposited on fused silica patterned with glassy carbon electrodes (100 μm spacings), and of the same film after conversion to $Cs_2AgBiI_6$ by anion exchange using TMSI. **(E)** Room-temperature photoconductivity action spectra (solid lines) and absorption spectra (dashed lines) of the same thermally evaporated film on carbon electrodes before ($Cs_2AgBiBr_6$, black) and after ($Cs_2AgBiI_6$, red) anion exchange with TMSI. Film thickness is ~200 nm. Excitation power densities were ~70 μW/cm$^2$. **(F)** Photocurrent as a function of incident optical power under 532 nm (2.33 eV) illumination with an illumination area of ~1.33 cm$^2$. The red dashed line represents a power-law fit ($I_{ph} \propto P^{\alpha}$), with α = 0.47. **(G)** Photocurrent switching response of a $Cs_2AgBiI_6$ device under periodic 532 nm illumination. All photoconductivity data were collected using an applied DC bias of 20 mV.

Photoconductivity measurements were performed with several such $Cs_2AgBiI_6$ devices by exciting the sample with the tunable output of a lamp + monochromator source. The discussion below focuses on devices with carbon electrodes, and results from devices with gold electrodes are presented in the Supporting Information (**Figures S24–26**). **Figure 7E** plots the room-temperature photoconductivity action spectrum of a representative $Cs_2AgBiI_6$ film in comparison with the absorption spectrum of the same film. From the film thickness of ca. 200 nm, the absorbance at the first feature near 2.4 eV corresponds to an absorption coefficient of $\alpha \approx 7\times10^4$ $cm^{-1}$, in good agreement with the value estimated from the molar extinction coefficient of $Cs_2AgBiI_6$ nanocrystals at this energy ($\alpha \approx 8.7 \times 10^4$ $cm^{-1}$);[34] these absorption coefficients are comparable to those of lead-halide perovskites. The photoconductivity data are plotted as responsivity, defined as $R = \frac{I_{\text{ph}}}{P}$, where $I_{\text{ph}}$ is the photocurrent and $P$ is the incident optical power. The responsivity shows a pronounced onset at the absorption edge and continues to higher energies. A weak tailing response is observed down to at least ~1.4 eV, consistent with sub-gap states contributing to the photoresponse. For reference, we also show the photoconductivity action spectrum and absorption spectrum of the same ($Cs_2AgBiBr_6$) film measured prior to the anion-exchange reaction. This comparison minimizes variations arising from film morphology or thickness, isolating the effect of the halide composition. The $Cs_2AgBiBr_6$ film also shows its photoconductivity onset at the same energy as its absorption onset, but its responsivity is generally ca. 3× lower than the $Cs_2AgBiI_6$ film's responsivity under the same measurement conditions, in addition to having a higher-energy onset. On gold electrodes, the contrast is even greater (**Figure S24**). This difference may reflect a more dispersive band structure and higher carrier mobility in the iodide elpasolite, consistent with computational results for this composition.[36] Dark currents are greater in the $Cs_2AgBiI_6$ films (**Figure S27B**), suggesting that higher defect densities may also contribute to this greater photoresponsivity.

Current–voltage characteristics measured under dark conditions are linear and symmetric, confirming ohmic contact between the $Cs_2AgBiI_6$ films and the glassy carbon or gold electrodes, while the photocurrents scale approximately linearly with applied bias over this range, consistent with field-driven photocarrier transport (**Figure S27**). The dependence of the photocurrent ($I_{\text{ph}}$) on incident optical power ($P$) at 2.33 eV follows a power-law relationship ($I_{\text{ph}} \propto P^{\alpha}$) with $\alpha \approx 0.47$ when using carbon contacts and $\alpha \approx 0.95$ when using gold contacts, where $\alpha = 1$ corresponds to the ideal trap-free photodetector. Both device types show good stability under chopped photoexcitation (**Figures 7G and S24**), but additionally show evidence of field-induced ion migration. Overall, these results demonstrate promising photoconductivity in $Cs_2AgBiI_6$. To our knowledge, these are the first such results for any 3D iodide double perovskite. Further work to address defects and grain boundaries in these films can be anticipated to improve upon these promising results for $Cs_2AgBiI_6$, along the lines of what has been achieved with other metal-halide perovskite semiconductors.

## Conclusion

Powders and films of the 3D iodide double perovskite, $Cs_2AgBiI_6$, have been prepared *via* anion exchange from phase-pure bulk $Cs_2AgBiBr_6$. We find that bulk $Cs_2AgBiI_6$ is kinetically stable in the absence of water, showing no detectable decomposition for at least five months at room temperature. We also find that strictly anhydrous processing conditions are necessary for complete bromide-to-iodide anion exchange within the elpasolite framework. Although generally considered air stable, trace water in the $Cs_2AgBiBr_6$ precursor adsorbed from air is identified as a critical factor inhibiting complete anion exchange. Such water is proposed to catalyze decomposition of $Cs_2AgBiI_6$ by increasing cation mobility and thereby accelerating the massive structural rearrangements that accompany decomposition. This water-catalyzed decomposition provides a mechanistic explanation for the prior elusiveness of bulk $Cs_2AgBiI_6$.[34–37]

Bulk $Cs_2AgBiI_6$ shows a narrow optical bandgap of $1.70 \pm 0.05$ eV, broad visible absorption, near-infrared photoluminescence, a surprising stability of at least months under inert conditions at room temperature, and thermal stability up to ~70 °C. The spontaneous decomposition of $Cs_2AgBiI_6$ above ~70 °C likely explains why this material was not obtained *via* thermal solid-state reaction between $Cs_3Bi_2I_9$ and $CsAg_2I_3$,[39] and this observation may help guide future solid-state syntheses of this material. Phase-pure $Cs_2AgBiI_6$ films prepared by $Cs_2AgBiBr_6$ evaporation followed by anion exchange display promising photoconductivity with a response that traces the compound's absorbance throughout the visible spectral range, demonstrating promising photocarrier generation and transport. Extending the morphology of this material from the nanoscale to its bulk polycrystalline and film forms opens a path toward environmentally benign metal-halide absorber materials for photovoltaic and optoelectronic applications and motivates the exploration of other yet-undiscovered bulk iodide elpasolite semiconductors that may be accessed *via* similar anion-exchange strategies.

**Supporting Information.** This material is available free of charge *via* the Internet at http://pubs.acs.org and includes: Detailed experimental procedures; Goldschmidt tolerance factor calculations; powder X-ray diffraction (XRD) data; Rietveld refinement of powder X-ray diffraction data; scanning electron microscopy (SEM) images; energy-dispersive X-ray (EDX) elemental maps and quantitative analysis; Fourier transform infrared (FTIR) spectroscopy; Raman spectra; additional photographs of powders, films and fabricated devices; absorption and diffuse reflectance spectra; photoluminescence (PL) spectra and their decay fitting parameters, photoconductivity spectra; and dark current–voltage characteristics.
Video S1. Representative time-lapse video showing the vapor-phase anion exchange of $Cs_2AgBiBr_6$ to $Cs_2AgBiI_6$. $Cs_2AgBiBr_6$ powder and evaporated thin-film samples were sealed in 20 mL vials together with separate vials containing trimethylsilyl iodide (TMSI). The video captures the first 3 h of the reaction and highlights the pronounced color change associated with iodide incorporation and formation of $Cs_2AgBiI_6$.

**Data availability**
The original data presented in this work will be made freely available at zenodo.org upon manuscript publication.

**Notes.** The authors declare no competing financial interest.

**Acknowledgments.** This work was supported by the U.S. National Science Foundation, Division of Materials Research (solid state and materials chemistry) through award DMR-2404703. The authors gratefully acknowledge postdoctoral fellowship support for F. H. from the Washington Research Foundation (WRF). Additional postdoctoral support from the Fulbright U.S. Scholar Program, Zuckerman STEM Leadership Program, Israel Scholarship Education Foundation (ISEF), and Israel Council for Higher Education (CHE) fellowships during early stages of this work is gratefully acknowledged. N.J.A. thanks the University of Washington Clean Energy Institute for graduate fellowship support. C.R.-M. gratefully acknowledges support from an NSF Graduate Research Fellowship (DGE-2140004). Part of this work was conducted at the Molecular Analysis Facility, a National Nanotechnology Coordinated Infrastructure (NNCI) site at the University of Washington that is supported in part by the National Science Foundation *via* awards NNCI-1542101 and NNCI-2025489, the U.W. Molecular Engineering & Sciences Institute, and the U.W. Clean Energy Institute. SEM and EDX experiments were performed using the University of Washington MEM-C Shared Facilities, supported by the U.S. National Science Foundation *via* Materials Research Science and Engineering Center (MRSEC) award DMR-2308979. The authors express their gratitude to Dr. Samantha Young for her assistance with the XPS measurements and to Todd Lewis for guidance on the fabrication of the glassy carbon electrodes used in this study.

## Table of Contents artwork

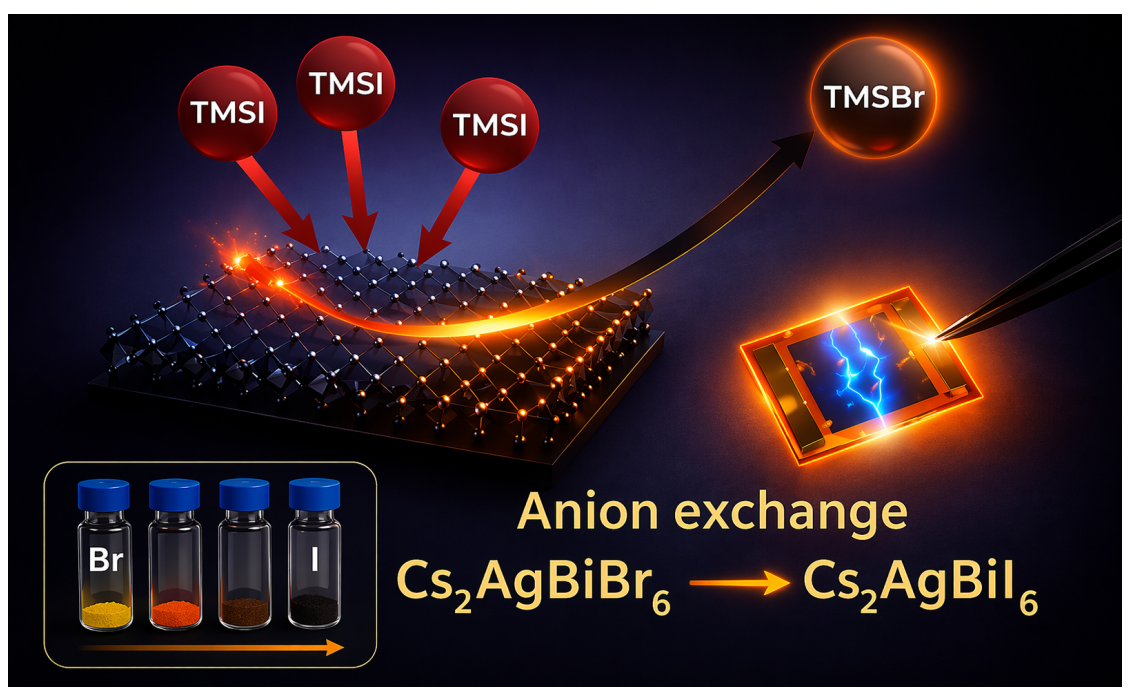

*Supporting Information for:*

# Synthesis of the Elusive Bulk Iodide Double-Perovskite Semiconductor, $Cs_2AgBiI_6$: Microcrystals and Photoconductive Films

Faris Horani, Nicolas Nguyen, Carmelita Ro-Mendez, Nicholas J. Adams, and Daniel R. Gamelin*

*Department of Chemistry, University of Washington, Seattle, Washington 98195-1700, United States*
*Email: gamelin@uw.edu

## I. Experimental Methods

**Materials.** Cesium bromide (CsBr, 99.999%) and trimethylsilyl iodide (TMSI) were purchased from Sigma-Aldrich. Silver bromide (AgBr, 99.9%) and bismuth(III) bromide ($BiBr_3$, 97%) were obtained from Strem Chemicals. All chemicals were used as received without further purification.

**Synthesis of $Cs_2AgBiBr_6$ powder.** Anhydrous CsBr (4.0 mmol), AgBr (2.0 mmol), and $BiBr_3$ (2.0 mmol) were combined with toluene (10 mL) in a 50 mL yttria-stabilized zirconia (YSZ) milling jar, together with ~16 g of 6 mm diameter YSZ milling balls. The sealed jar was loaded into a planetary ball mill (MSE Supplies; PMGB-0.2L) and milled for 8 h at 500–700 rpm. Unless otherwise specified, the resulting yellow powder was isolated by vacuum filtration and dried, then transferred into a glovebox and annealed at 250 °C for 30 min to react residual AgBr, yielding phase-pure $Cs_2AgBiBr_6$.

**Synthesis of $Cs_2AgBiI_6$ powder *via* anion exchange.** $Cs_2AgBiI_6$ powder was obtained *via* post-synthetic anion exchange of $Cs_2AgBiBr_6$ using trimethylsilyl iodide (TMSI) vapor. In a typical procedure, 50–100 mg of $Cs_2AgBiBr_6$ and a separate 1.5 mL vial containing 0.5 mL of neat TMSI were placed inside an inert, sealed 20 mL reaction vial. The starting powder was dark yellow and progressively converted to a black solid over 24 h. After completion of the exchange, the TMSI source vial was removed and the product was placed under dynamic vacuum to eliminate residual TMSBr or TMSI. The resulting $Cs_2AgBiI_6$ powder was stored in an inert-atmosphere glovebox and showed no sign of decomposition after several months.

**Thermal evaporation of $Cs_2AgBiBr_6$ thin films.** $Cs_2AgBiBr_6$ thin films were prepared by single-source vapor deposition as described previously.[1,2] Briefly, phase pure $Cs_2AgBiBr_6$ powder was loaded onto a tantalum evaporation boat. Soda-lime glass and silicon substrates were cleaned sequentially with hexane, ethanol, and acetone and dried in a 120 °C oven overnight. Substrates were mounted side-by-side 15 cm above the evaporation boat. The chamber was evacuated to ~1.5 x $10^{-5}$ Torr, and then the powder was sublimated by passing a high current through the evaporation boat. Approximately 75 mg was evaporated for optical-quality films and 200 mg was used for thicker films for X-ray diffraction. The samples were immediately transferred *via* ambient air to a nitrogen-filled glove box and annealed for 30 min at 250 °C on a hot plate.

**Synthesis of $Cs_2AgBiI_6$ thin films by anion exchange.** Halide exchange from bromide to iodide was carried out on evaporated $Cs_2AgBiBr_6$ thin films using trimethylsilyl iodide (TMSI) vapor. For a typical exchange reaction, a $Cs_2AgBiBr_6$ film and a separate 1.5 mL vial containing 0.5 mL of neat TMSI were enclosed together in a tightly sealed 20 mL vial under nitrogen. Exposure to TMSI vapor induced a gradual color change of the film from dark yellow to dark red over 24 h, consistent with bromide-to-iodide conversion. Following the exchange, the $Cs_2AgBiI_6$ films were removed and placed under dynamic vacuum to eliminate residual TMSBr or TMSI. The final $Cs_2AgBiI_6$ films were stored under inert atmosphere for further characterization.

**Ambient stability and moisture-sensitivity tests.** To evaluate the relative sensitivity of $Cs_2AgBiI_6$ to moisture vs dry air, the powder was placed in a loosely capped vial (to allow exposure to the ambient atmosphere) and transferred into a 20 mL container containing calcium sulfate desiccant (Drierite). The outer container was left open to ambient atmosphere for 1 h to allow equilibration prior to sealing. The sample was then stored under these conditions and monitored over time. No discernible color change was observed over 3 weeks, with the material retaining its characteristic black appearance. Powder X-ray diffraction (XRD) analysis following storage for 3 weeks confirmed preservation of the phase-pure elpasolite structure (Figure S11).

**FTIR measurements.** Fourier transform infrared (FTIR) spectra were acquired on a PerkinElmer spectrometer in attenuated total reflectance (ATR) mode with a spectral resolution of 2 $cm^{-1}$. For $N_2$-annealed $Cs_2AgBiBr_6$ powders, a Fluorolube mull was prepared to minimize exposure to ambient moisture during sample transfer and measurement.

**X-ray diffraction (XRD) measurements.** Benchtop powder XRD data were collected using a Bruker D8 Discover diffractometer equipped with a high-efficiency IμS microfocus Cu Kα X-ray source. Routine measurements were performed using a Bruker PILATUS 100K area detector. High-resolution pXRD data used for Rietveld refinement and quantitative impurity-phase analysis were collected using a Bruker LynxEye-XE-T 1D detector with a 0.2 mm receiving slit and an acquisition time of approximately 7 h to maximize sensitivity toward weak reflections. Powder samples were mounted on glass substrates and sealed with Kapton tape during data collection to minimize exposure to ambient moisture and oxygen.

**Electron microscopy.** Scanning electron microscopy (SEM) images were acquired using a Phenom Pharos G2 Desktop field-emission SEM operating at accelerating voltages of 10–20 kV. SEM samples were prepared inside a glovebox by mounting powders or films onto conductive double-sided carbon tape (Electron Microscopy Sciences). Energy-dispersive X-ray spectroscopy (EDS) measurements were performed using a standard EDS detector.

**Raman scattering measurements.** Raman measurements were performed using a Renishaw inVia Raman microscope equipped with a 1200 grooves/mm grating and a Leica DMIRBE inverted optical microscope. A 785 nm laser was used as the excitation source. The collimated monochromatic beam was focused onto the sample through a Leica objective lens. Raman scattered light was collected by the same objective, directed into the inVia spectrometer, dispersed by the diffraction grating, and focused onto a CCD camera.

**X-ray photoelectron spectroscopy (XPS) measurements.** XPS data were collected on a Kratos Axis Supra+ spectrometer. This instrument uses a monochromatized Al $K_\alpha$ X-ray source and a low-energy electron flood gun for charge neutralization. The X-ray spot size for these acquisitions was on the order of 700 x 300 µm. Pressure in the analytical chamber during spectral acquisition was less than $5 \times 10^{-9}$ Torr. Pass energy for survey and detailed spectra was 160 eV. Pass energy for high-resolution spectra was 20 eV. The take-off angle was 0º (100 Å sampling depth). The Kratos ESCApe software was used to determine peak areas and to fit peaks in the high-resolution spectra. For the high-resolution spectra, a Shirley background was used, and all binding energies were referenced to the C ls C-C bonds at 285.0 eV.

**Diffuse reflectance UV–Vis–NIR spectroscopy.** Diffuse reflectance data were recorded using a Cary 5000 UV-Vis-NIR Spectrophotometer, with $BaSO_4$ (barium sulfate) used as the reference standard.

**Photoluminescence (PL) measurements.** $Cs_2AgBiI_6$ powder was sandwiched between two sapphire disks and mounted in a closed-cycle helium cryostat under high vacuum ($10^{-6}$ Pa). Samples were excited using a continuous-wave 405 nm (3.06 eV) diode. Spectra were recorded using a liquid-$N_2$-cooled silicon CCD camera and a liquid-$N_2$-cooled NIR PMT. All PL spectra have been corrected for instrument response. All luminescence spectra are displayed as photon counts vs energy (eV) following appropriate conversion.[3,4]

**Photoconductivity measurements.** Thin-film samples for photoconductivity measurements were prepared by depositing $Cs_2AgBiBr_6$ films directly onto commercial gold interdigitated electrodes with a finger spacing of 20 µm (Newvision1981) or glassy carbon electrodes with a finger spacing of 100 µm. Glassy carbon electrodes were fabricated by pyrolysis of patterned SU-8 photoresist on fused silica substrates. The electrode geometry was defined by contact photolithography, followed by development and pyrolysis under flowing $N_2$ at 1000 °C for 1 h in a tube furnace to convert the patterned photoresist into conductive glassy carbon electrodes. Pairs of thickness-matched films were prepared by simultaneous evaporation onto side-by-side substrates. The as-deposited films were annealed at 250 ºC in a glovebox for 1 h. After annealing, $Cs_2AgBiBr_6$ samples selected for conversion to the iodide composition (*e.g.*, one of a matched pair) were placed in sealed containers and exposed to TMSI vapor for 48 h, as detailed above. Three devices were made for each electrode type (glassy carbon and gold), and the results were found to be highly reproducible for each set.

Air-free photoconductivity measurements were performed on the completed devices under dry nitrogen. A Keithley 2400 SMU was used to apply a DC bias across the interdigitated electrodes. The resulting photocurrent was detected using a Keithley 427 current preamplifier and a Stanford Research Systems SR830 lock-in amplifier. The excitation source consisted of a halogen lamp, optically chopped at ~30 Hz and spectrally filtered using a double monochromator. The resulting illumination was incident on the device with a power density of ~5 $\mu W/cm^2$, tunable over the 450–1000 nm (2.76–1.24 eV) range, with a spectral resolution of ca. 10 nm. Higher-energy spectral features were independently verified using a xenon lamp coupled to a single monochromator, delivering ~70 $\mu W/cm^2$ of excitation with a spectral bandwidth of 20 nm.

Photoconductivity power-dependence and on/off photoresponse measurements were performed under 532 nm illumination from an Arroyo Instruments laser diode, modulated at 30 Hz by a Stanford Research Systems DS345 function generator. The modulated photocurrent was detected using a Keithley 427 current preamplifier and a Stanford Research Systems SR830 lock-in amplifier. Incident optical power was varied using a gradient neutral-density filter wheel and measured at the sample plane using a power meter. A DC bias of 20 mV was applied across the electrodes during measurement.

## II. Stabilities of Metal-Halide Double Perovskites (Elpasolites)

**Goldschmidt tolerance factor.** The structural stability of metal–halide perovskites is commonly evaluated using the Goldschmidt tolerance factor, *t*, which for conventional $AMX_3$ perovskites is defined as given in eq 1.[5]

$$t = \frac{R_A + R_X}{\sqrt{2}(R_M + R_X)} \tag{1}$$

Here, $R_A$, $R_M$, and $R_X$ denote the ionic radii of the A-site cation, B-site metal cation, and halide anion, respectively. For double perovskites with the general formula $A_2M^{(I)}M^{(III)}X_6$ (elpasolites), the crystal structure (space group *Fm*–3*m*) contains two chemically distinct $MX_6$ octahedra. To account for this, the effective B-site ionic radius is approximated by the arithmetic mean of the two metal cation radii:

$$\langle R_M \rangle = \frac{R_{M(I)} + R_{M(III)}}{2} \tag{2}$$

Substituting $\langle R_M \rangle$ for $R_M$ in eq 1 yields the modified tolerance factor for double perovskites:[6]

$$t = \frac{R_A + R_X}{\sqrt{2}(\langle R_M \rangle + R_X)} \tag{3}$$

Based on eq 3, the stability window for the formation of a three-dimensional double-perovskite (elpasolite) structure can be described by the following empirical conditions:

$$R_A > 0.73 R_X \tag{4}$$

$$0.41 R_X \leq \langle R_M \rangle \leq 0.86 R_A - 0.13 R_X \tag{5}$$

$$R_A > \langle R_M \rangle \tag{6}$$

In addition, the average octahedral factor must satisfy the constraint:

$$\langle R_M \rangle / R_X \geq 0.41 \tag{7}$$

Based on these criteria, the calculated tolerance factors and geometric parameters for the selected $A_2M^{(I)}M^{(III)}X_6$ compounds are summarized below.

| $A_2M^{(I)}M^{(III)}X_6$ | $R_A$ (Å) | $R_{M(I)}$ (Å) | $R_{M(III)}$ (Å) | $\langle R_M \rangle$ (Å) | $R_X$ (Å) | $t$ | $\langle R_M \rangle / R_X$ | eq (4) | eq (5) | eq (6) |
|---|---|---|---|---|---|---|---|---|---|---|
| $Cs_2AgBiBr_6$ | 1.88 | 1.15 | 1.03 | 1.09 | 1.96 | 0.89 | 0.55 | √ | √ | √ |
| $Cs_2AgBiI_6$ | 1.88 | 1.15 | 1.03 | 1.09 | 2.20 | 0.88 | 0.50 | √ | √ | √ |

**Modified tolerance factor.** A modification of the above tolerance factor was recently proposed by Bartel *et al.*[7] and is expressed as:

$$t' = \frac{R_X}{\langle R_M \rangle} - n_A\left(n_A - \frac{\frac{R_A}{\langle R_M \rangle}}{\ln\left(\frac{R_A}{\langle R_M \rangle}\right)}\right) \quad (8)$$

where $n_A$ is the A-site oxidation state. Double-perovskite structures are predicted to be stable when $t' < 4.18$. For $Cs_2AgBiI_6$ we calculate $t' = 4.18$, positioning this material precisely at the threshold of stability.

| $A_2M^{(I)}M^{(III)}X_6$ | $t'$ |
|---|---|
| $Cs_2AgBiBr_6$ | 3.96 |
| $Cs_2AgBiI_6$ | 4.18 |

### III. Additional Experimental Results

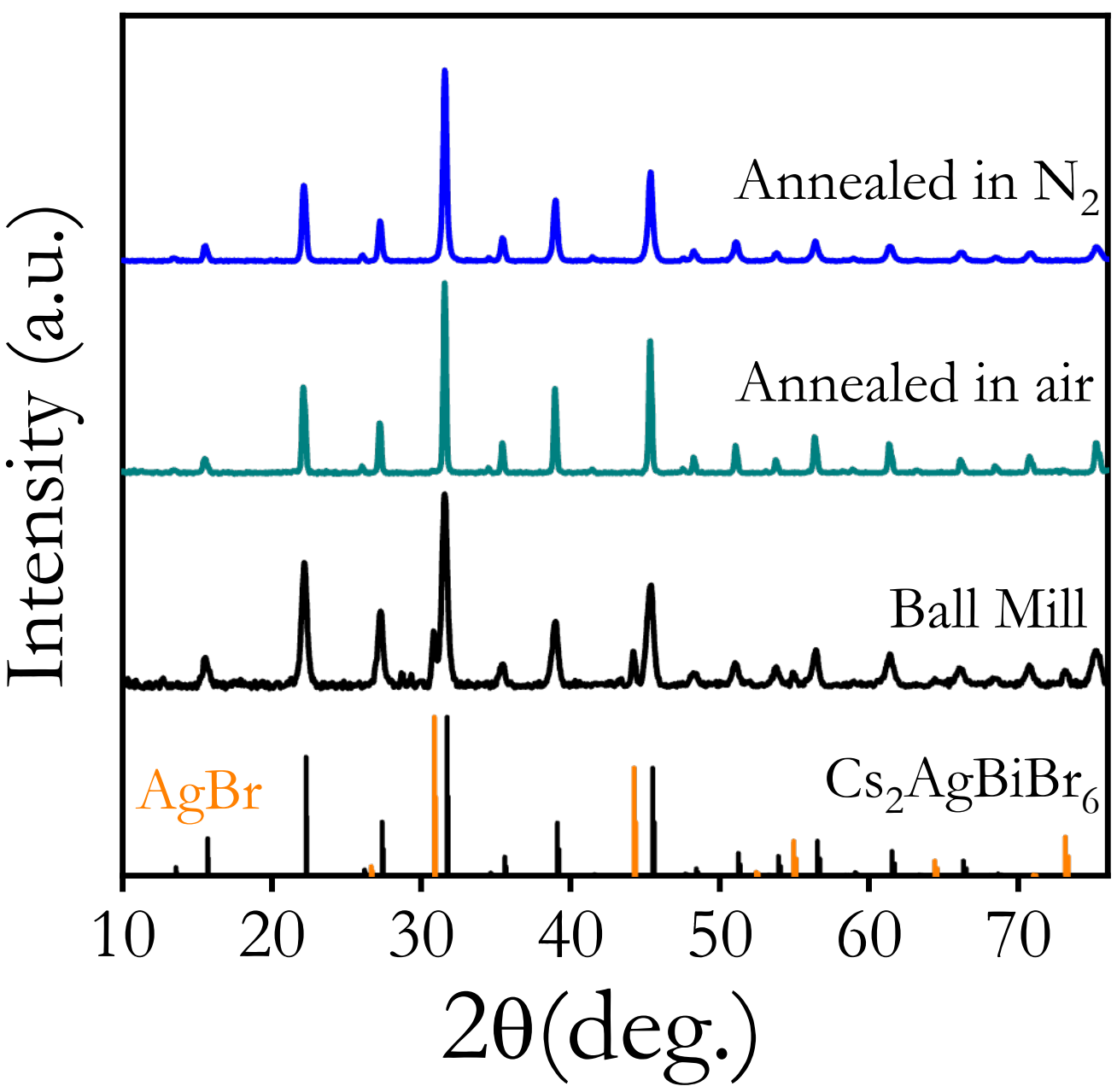

**Figure S1.** X-ray diffraction (XRD) data for ball-milled $Cs_2AgBiBr_6$ powder before annealing and after annealing at 250 °C. The as-milled sample shows additional reflections associated with secondary/binary phases (*e.g.*, AgBr), whereas annealing at 250 °C under either ambient air or $N_2$ eliminates these impurity phases and yields in the phase-pure $Cs_2AgBiBr_6$ elpasolite.

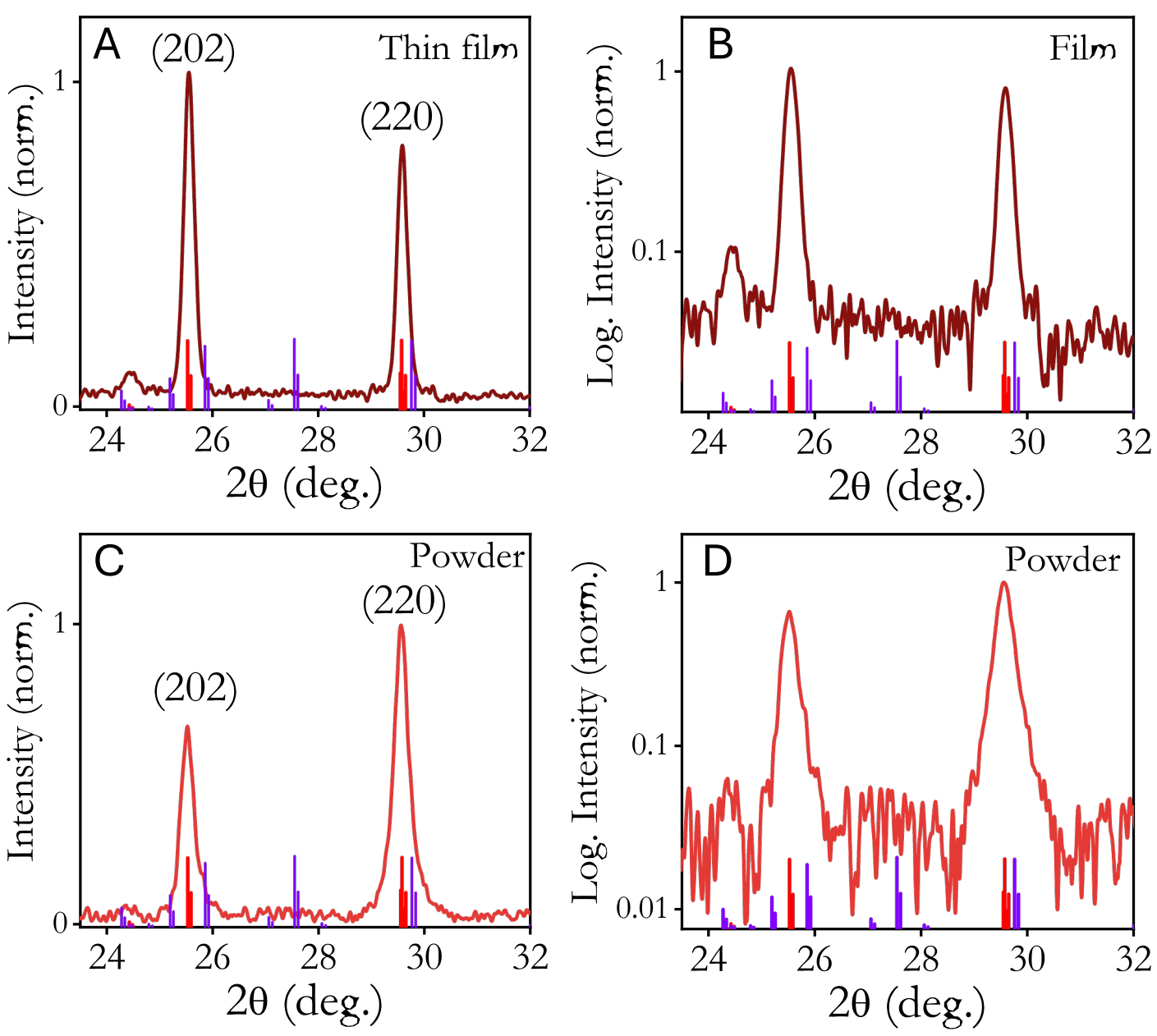

**Figure S2**. Powder X-ray diffraction data for microcrystalline $Cs_2AgBiI_6$ plotted on **(A)** linear and **(B)** logarithmic intensity scales. Powder X-ray diffraction data for thin-film $Cs_2AgBiI_6$ plotted on **(C)** linear and **(D)** logarithmic intensity scales. Reference patterns for $Cs_2AgBiI_6$ and $Cs_3Bi_2I_9$ are shown at the bottom of each panel in red and purple, respectively.

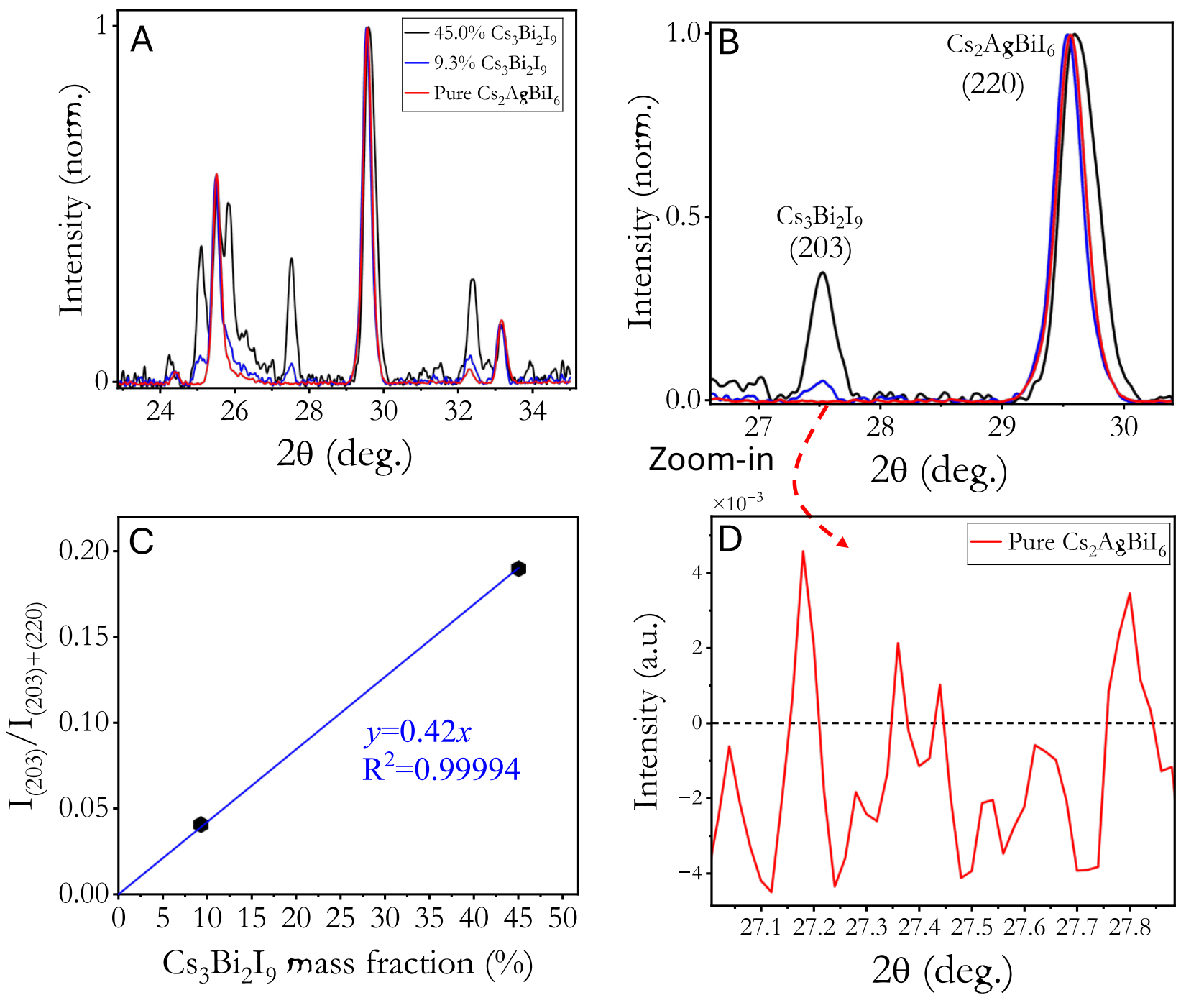

**Figure S3**. Estimation of the XRD impurity-detection threshold for $Cs_3Bi_2I_9$ in $Cs_2AgBiI_6$. **(A)** Powder X-ray diffraction patterns of phase-pure $Cs_2AgBiI_6$ and reference mixtures containing 9.3 wt% and 45.0 wt% $Cs_3Bi_2I_9$. **(B)** Expanded view of the diffraction region containing the characteristic $Cs_3Bi_2I_9$ (203) reflection and the neighboring $Cs_2AgBiI_6$ (220) reflection used for impurity analysis. **(C)** Calibration curve constructed from the normalized intensity ratio ($I_{(203)}/[I_{(203)}+I_{(220)}]$) as a function of $Cs_3Bi_2I_9$ mass fraction, yielding a linear relationship ($y = 0.42x$, $R^2 = 0.99994$). **(D)** Baseline fluctuations in a peak-free region adjacent to the $Cs_3Bi_2I_9$ (203) reflection. The standard deviation of the baseline noise was determined to be $\sigma \approx 0.0022$. Using the calibration curve in panel C and a 1σ criterion, the estimated impurity threshold corresponds to approximately 0.5 wt% $Cs_3Bi_2I_9$. Accordingly, the impurity content determined from laboratory X-ray diffraction measurements is estimated to have an uncertainty of approximately ±0.5 wt% $Cs_3Bi_2I_9$ near the detection threshold.

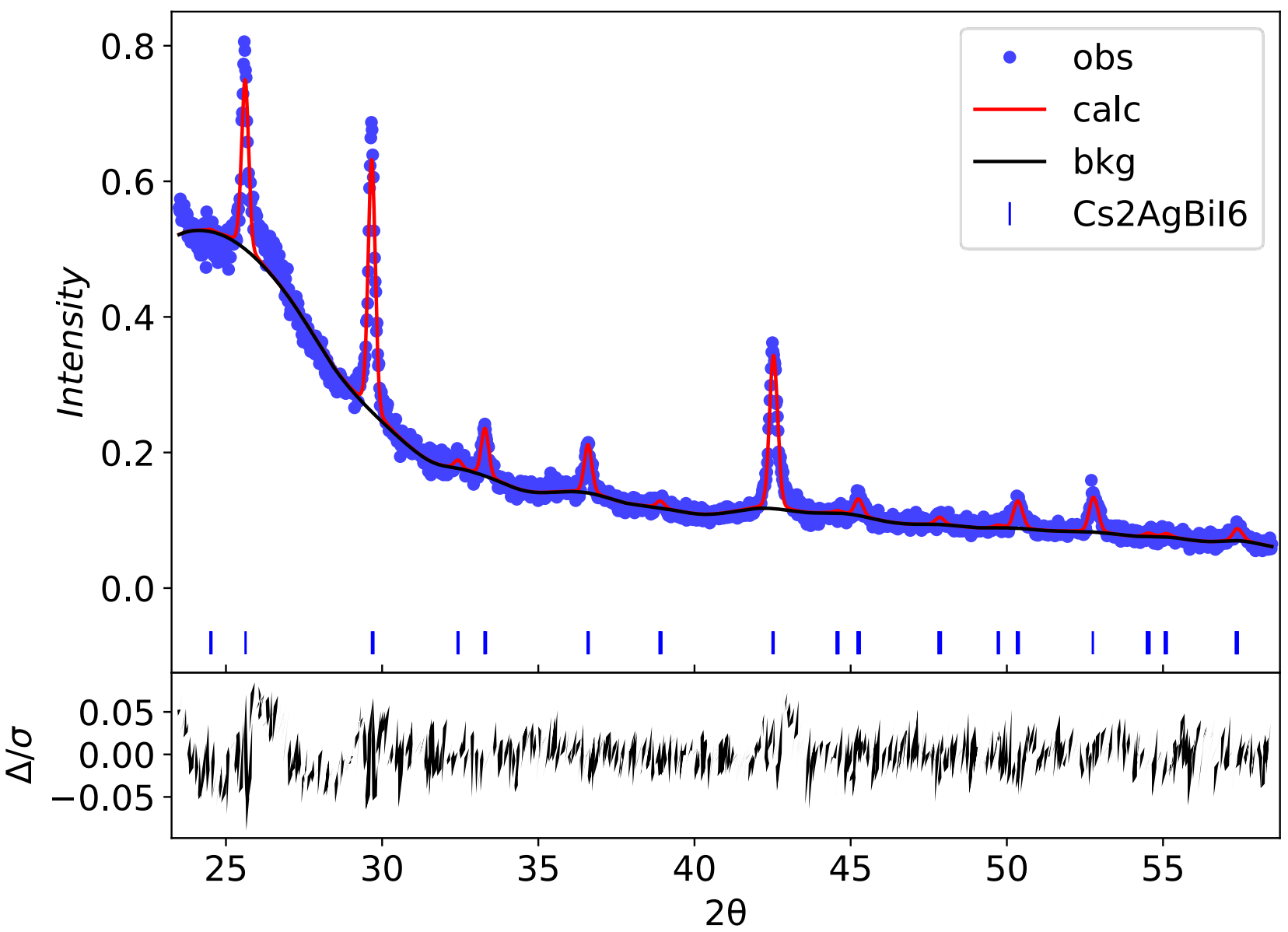

**Figure S4.** (A) Rietveld refinement of the powder X-ray diffraction data from bulk microcrystalline $Cs_2AgBiI_6$ using a tetragonal structural model. Blue circles represent experimental data, the red line corresponds to the calculated pattern, and the black line denotes the refined background. Vertical tick marks indicate the Bragg reflection positions of $Cs_2AgBiI_6$. The lower panel shows the difference profile ($I_{obs} - I_{calc}$). The refinement yields excellent agreement between the observed and calculated diffraction patterns, with $R_w$ = 5.75% and a goodness-of-fit parameter of $\chi^2 = 1.1$. The broad feature between 2θ ≈ 25° and 28° arises from the Kapton tape used to encapsulate the sample during data collection.

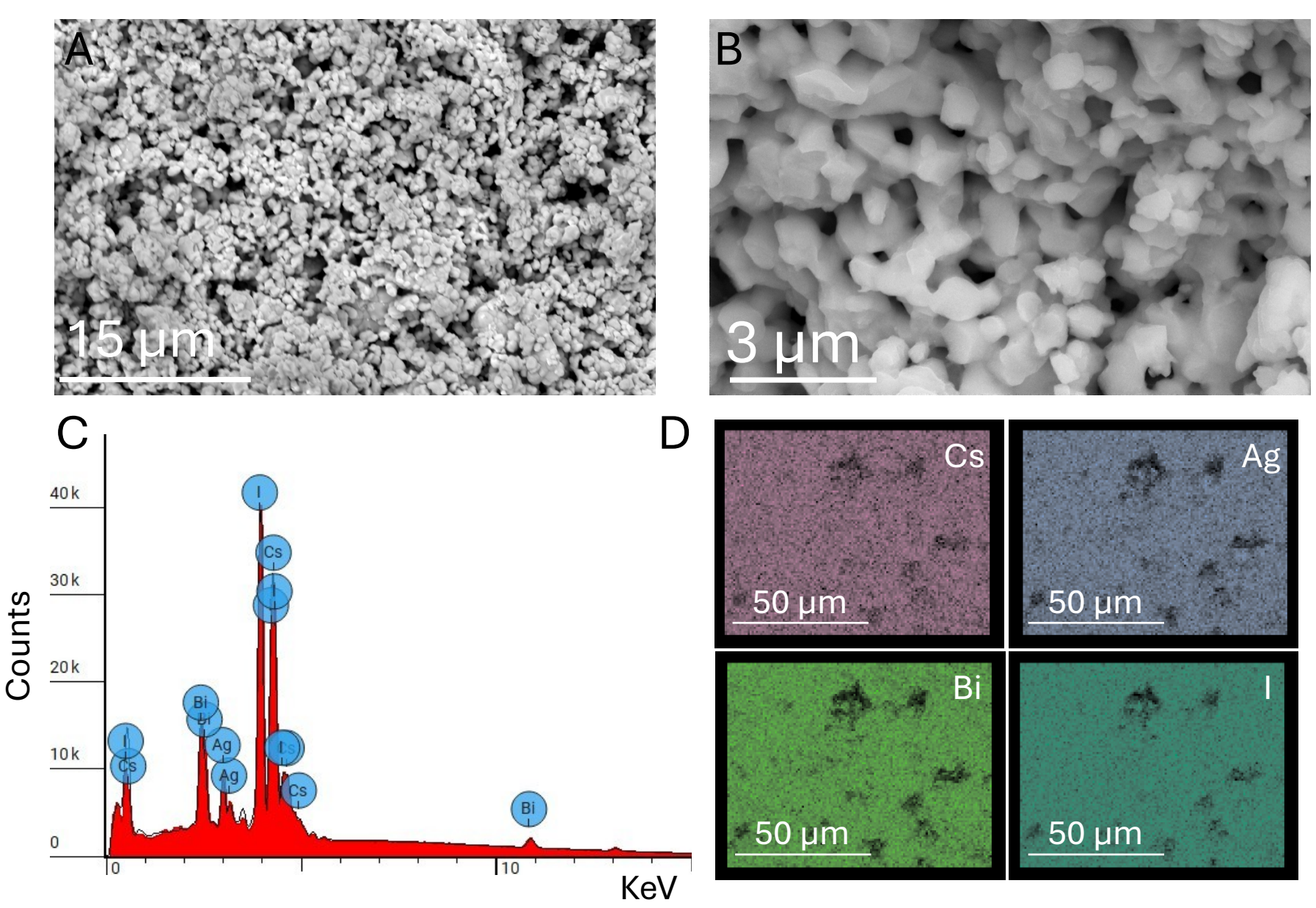

**Figure S5. (A, B)** SEM micrographs of $Cs_2AgBiI_6$ powder at different magnifications. **(C)** EDX spectrum of $Cs_2AgBiI_6$ powder. **(D)** Corresponding EDX elemental mapping showing a homogenous spatial distribution of Cs, Ag, Bi, and I.

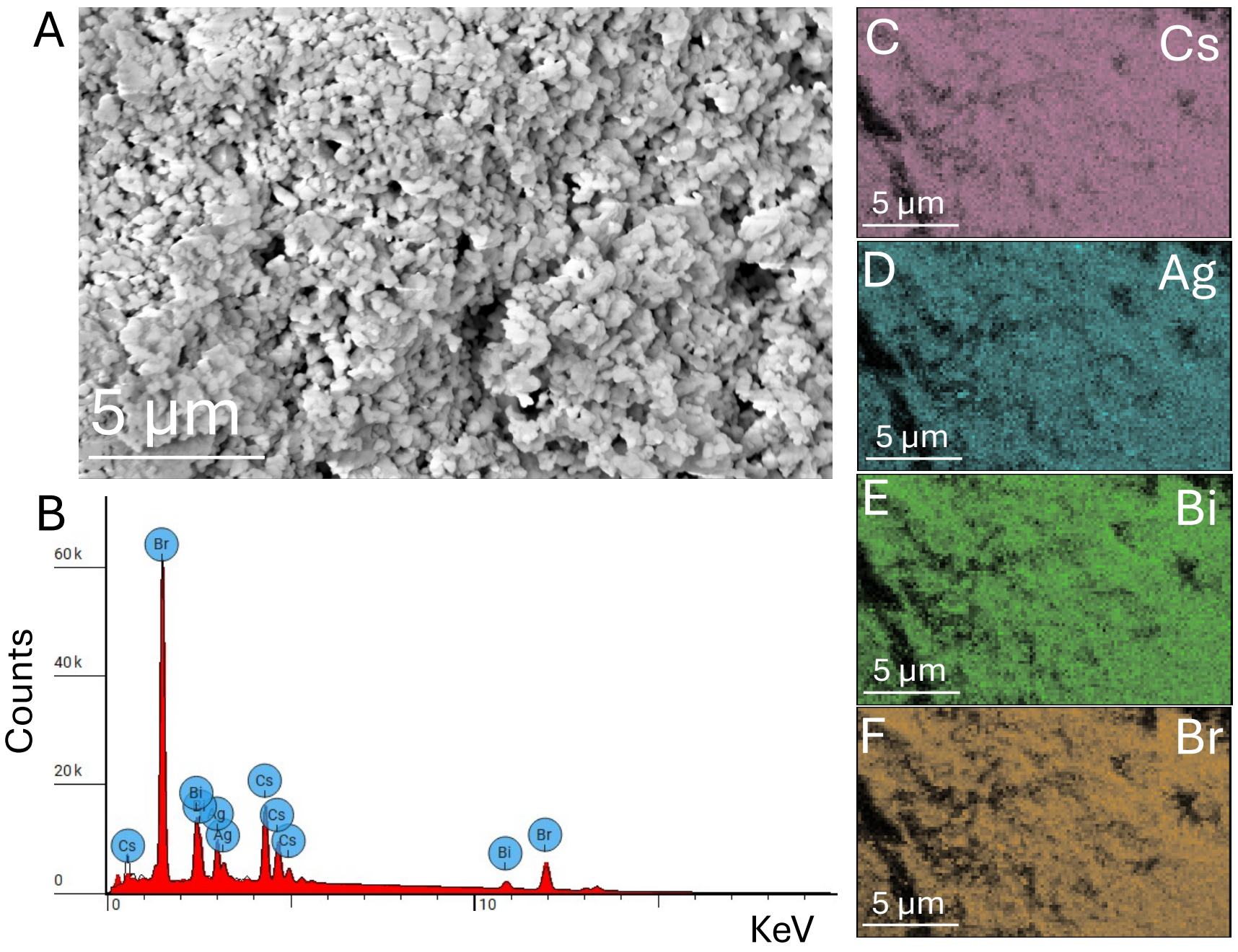

**Figure S6. (A)** SEM micrographs of $Cs_2AgBiBr_6$ powder annealed under $N_2$. **(B)** EDX spectrum of $Cs_2AgBiBr_6$ powder. **(C-F)** Corresponding EDX elemental mapping showing a homogenous spatial distribution of Cs, Ag, Bi, and Br.

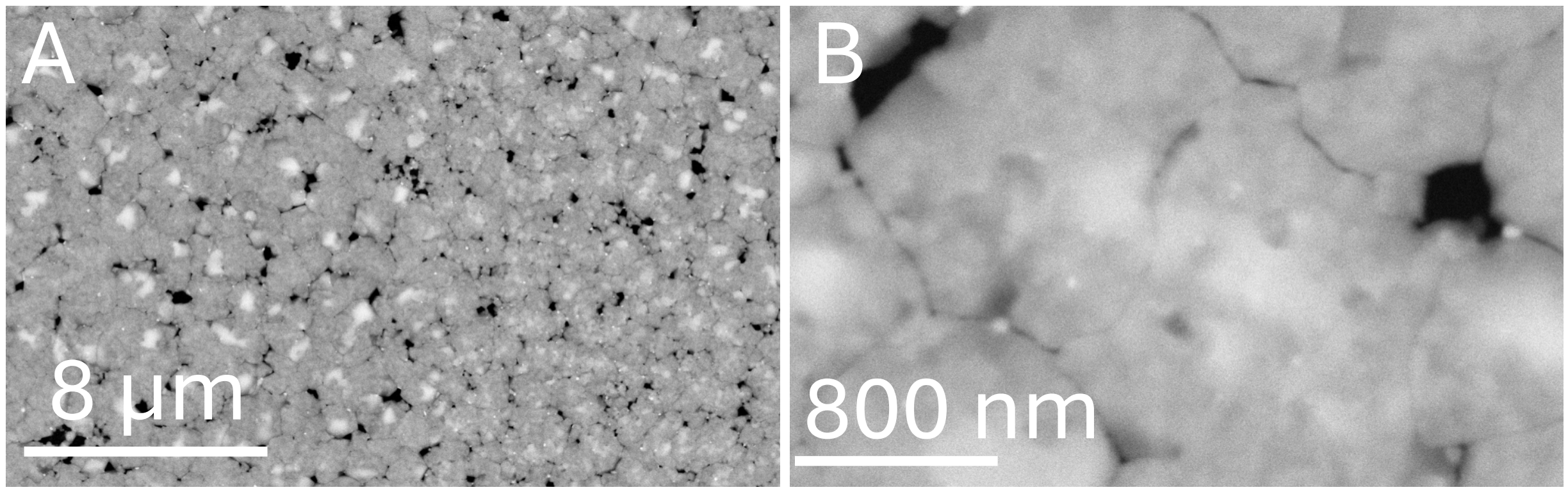

**Figure S7.** SEM images of the thermally evaporated $Cs_2AgBiBr_6$ film: **(A)** low-magnification overview and **(B)** high-magnification showing the grain structure. Bright spots are attributed to beam-induced damage.

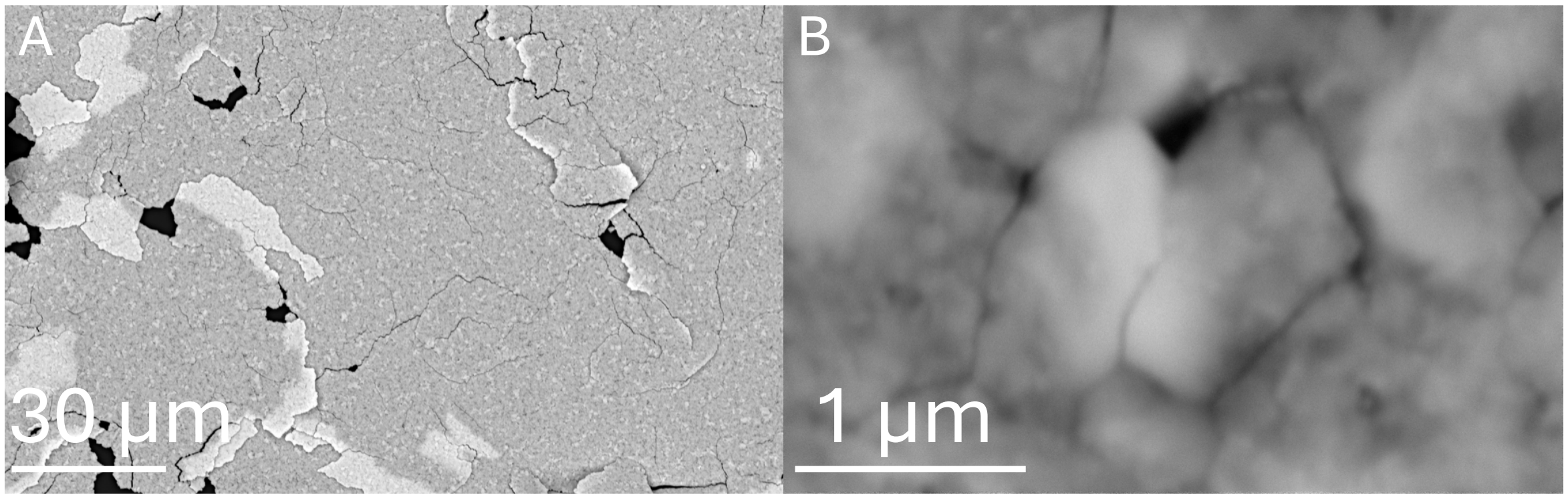

**Figure S8.** **(A)** Low-magnification and **(B)** high-magnification SEM images of a $Cs_2AgBiI_6$ thin film obtained by anion exchange of a thermally evaporated $Cs_2AgBiBr_6$ thin film *via* 24 h of exposure to TMSI vapor.

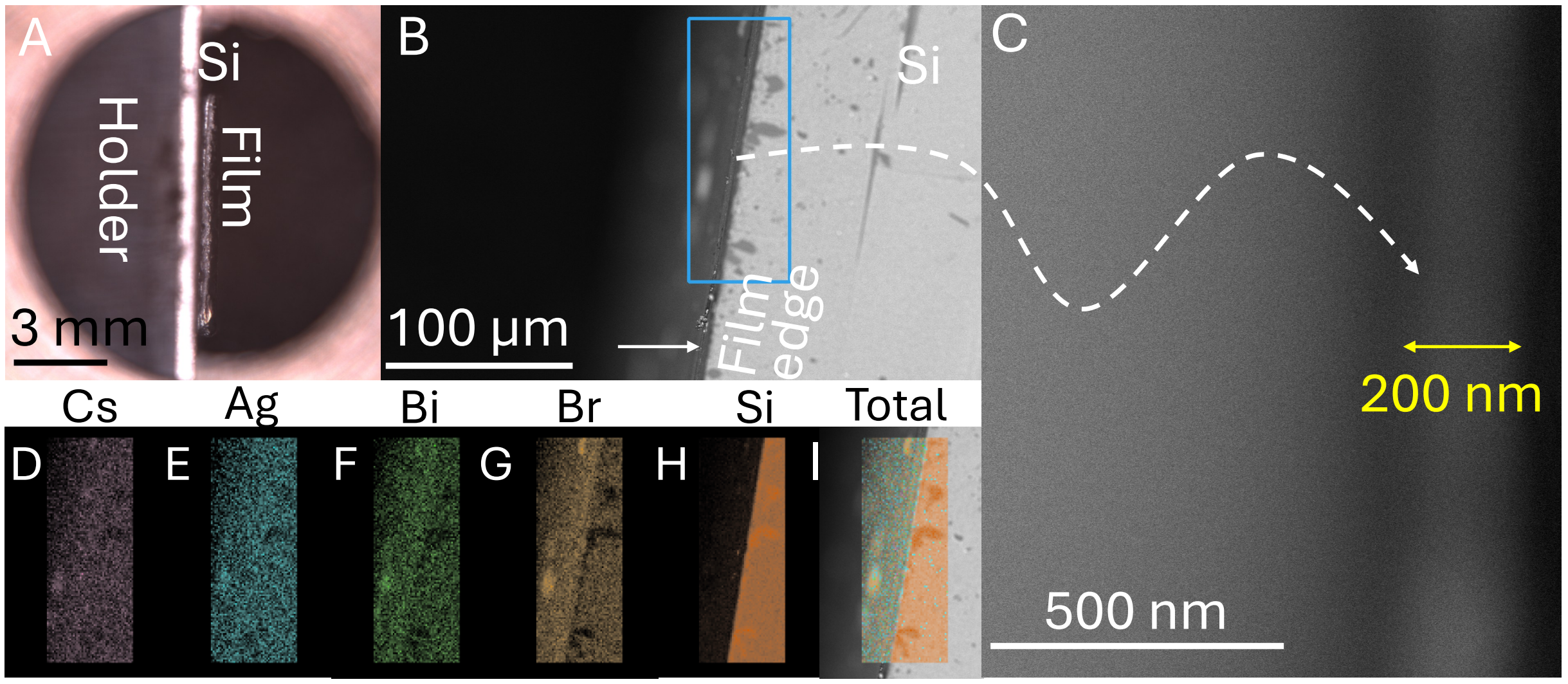

**Figure S9.** SEM images of a thermally evaporated $Cs_2AgBiBr_6$ film on a silicon substrate: **(A)** low-magnification overview and **(B, C)** magnified view of the film edge. **(D–I)** Corresponding EDX elemental maps of Cs, Ag, Bi, Br, and Si, acquired from the blue-framed region in panel (B).

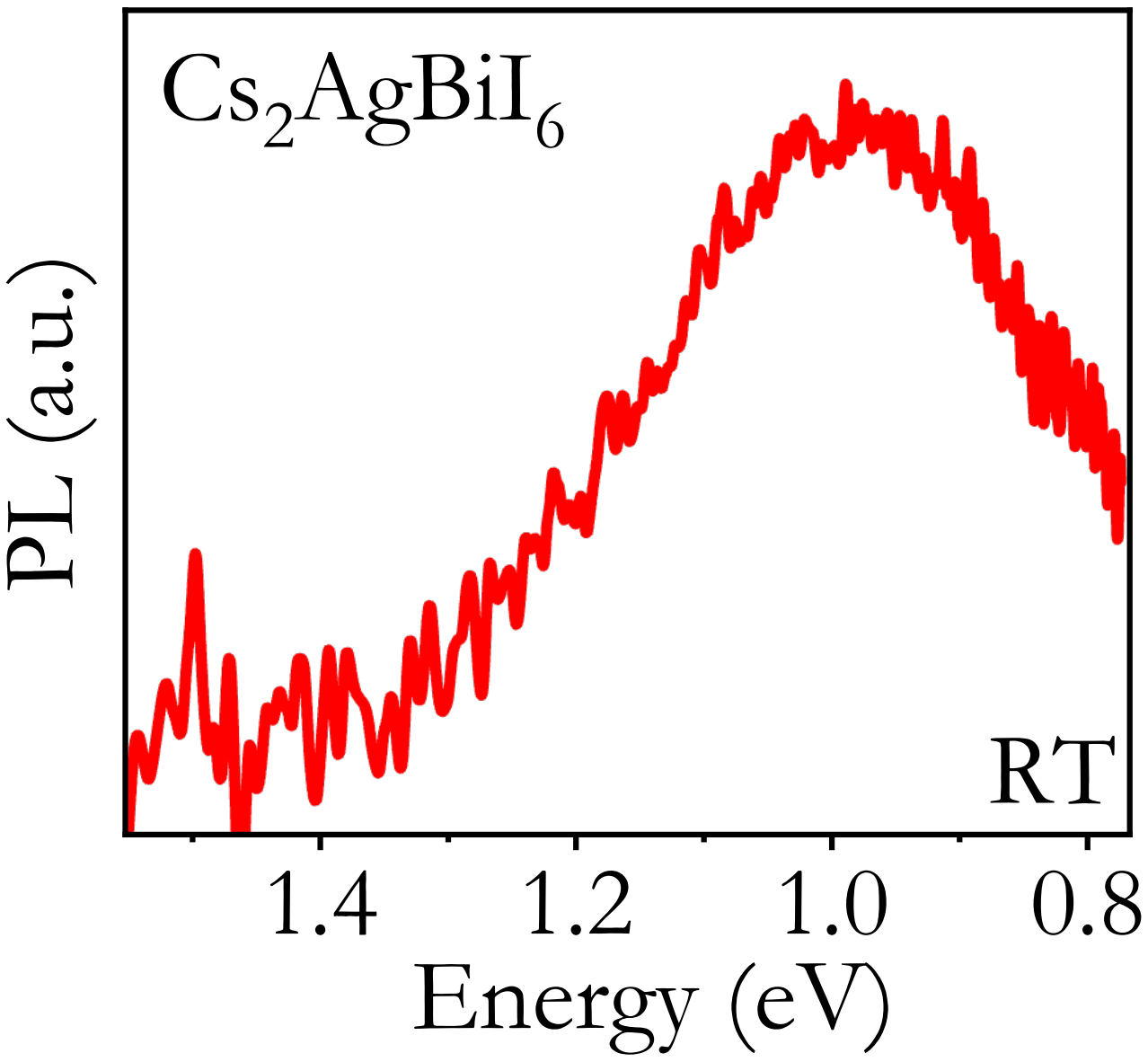

**Figure S10.** Photoluminescence (PL) spectrum of $Cs_2AgBiI_6$ measured at room temperature, showing weak emission in the near-infrared (NIR) region.

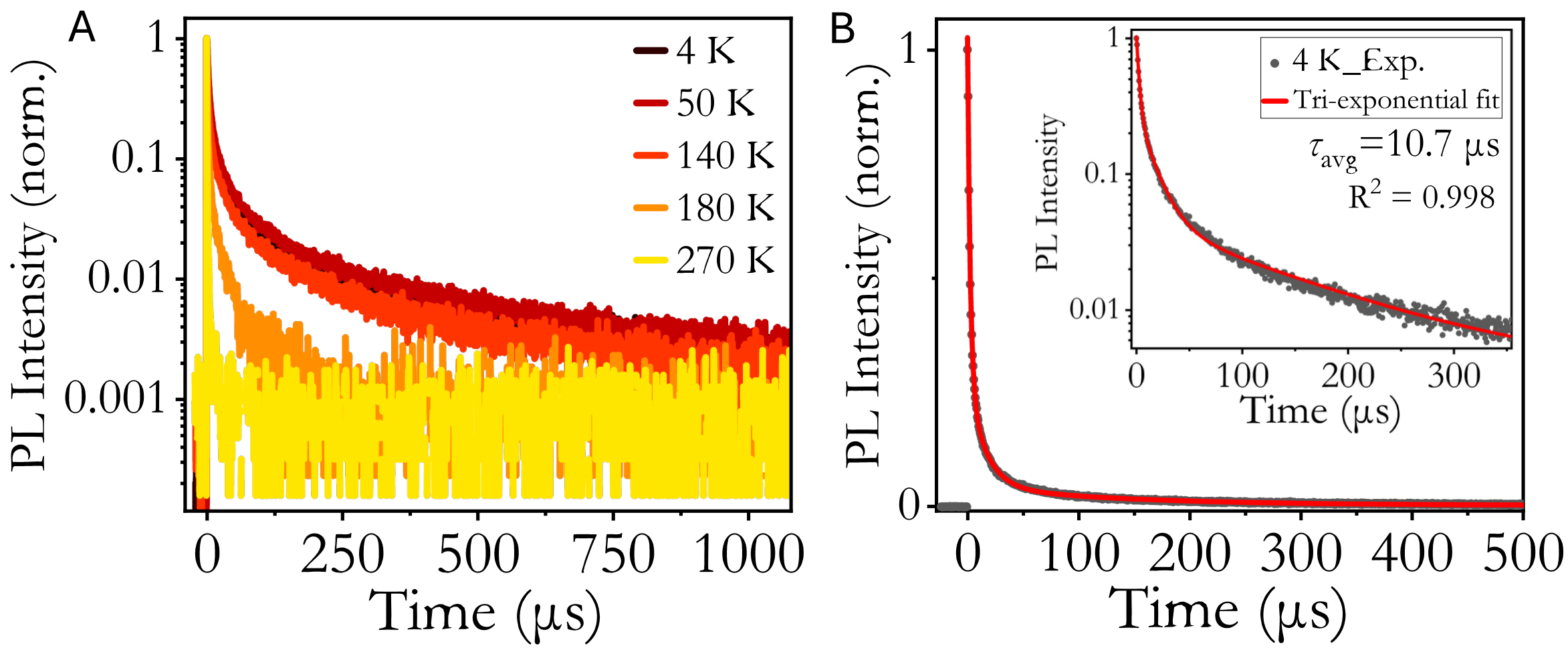

**Figure S11**. Variable-temperature photoluminescence decay data for $Cs_2AgBiI_6$ microcrystals excited with 355 nm laser pulses and monitored at 1.03 eV (1200 nm). **(A)** Normalized PL decay curves collected at different temperatures and displayed on a logarithmic intensity scale. **(B)** The 4 K PL decay curve shown on a linear scale together with the corresponding tri-exponential fit (red line). The inset shows the same data plotted on a logarithmic intensity scale. The average lifetime ($\tau_{avg}$) and goodness-of-fitting parameter ($R^2$) are indicated in the figure. The detailed fitting results for all temperatures are summarized in Table S2.

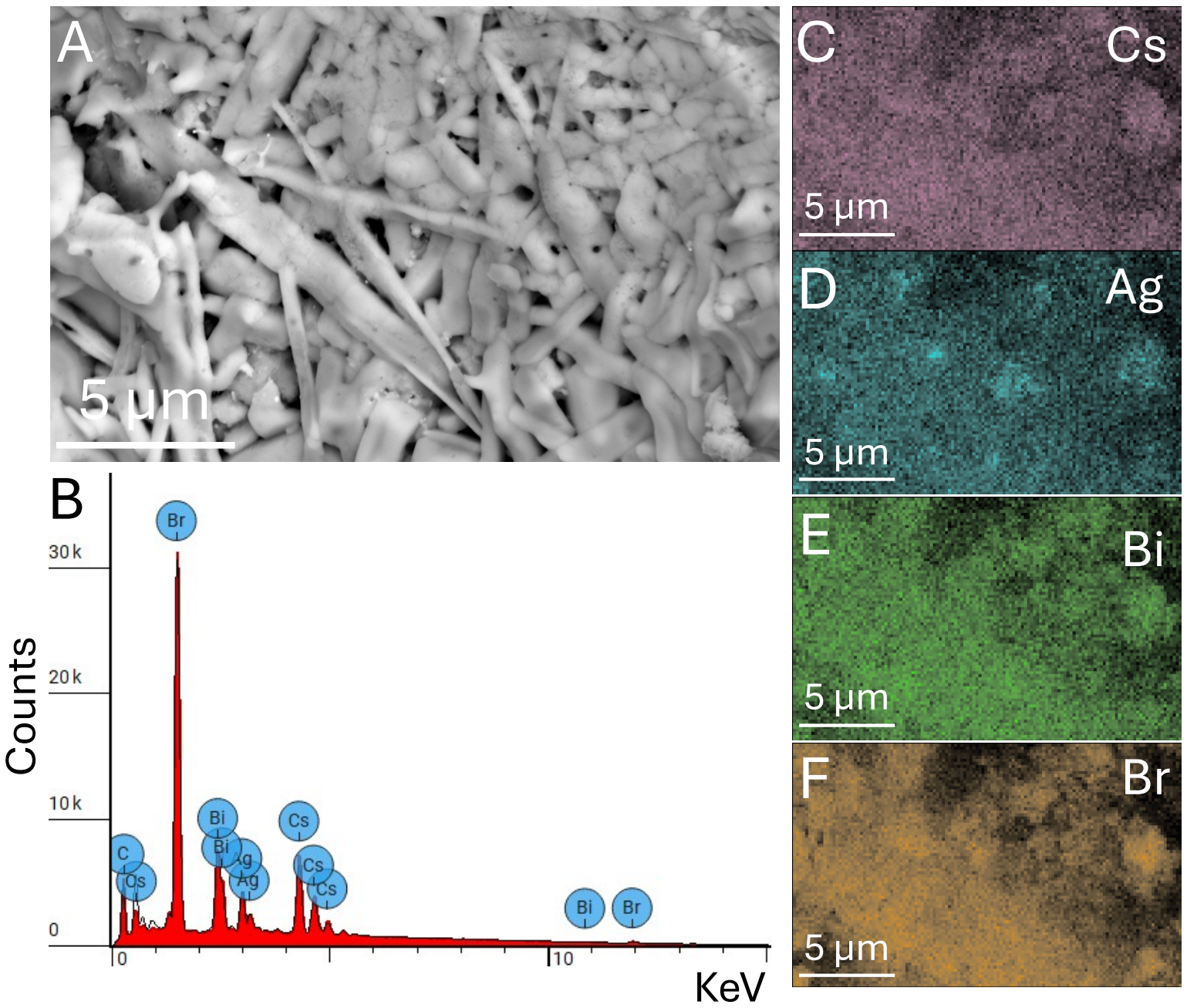

**Figure S12. (A)** SEM micrographs of $Cs_2AgBiBr_6$ powder annealed in ambient air. **(B)** EDX spectrum of $Cs_2AgBiBr_6$ powder. **(C-F)** Corresponding EDX elemental mapping showing the spatial distribution of Cs, Ag, Bi, and Br.

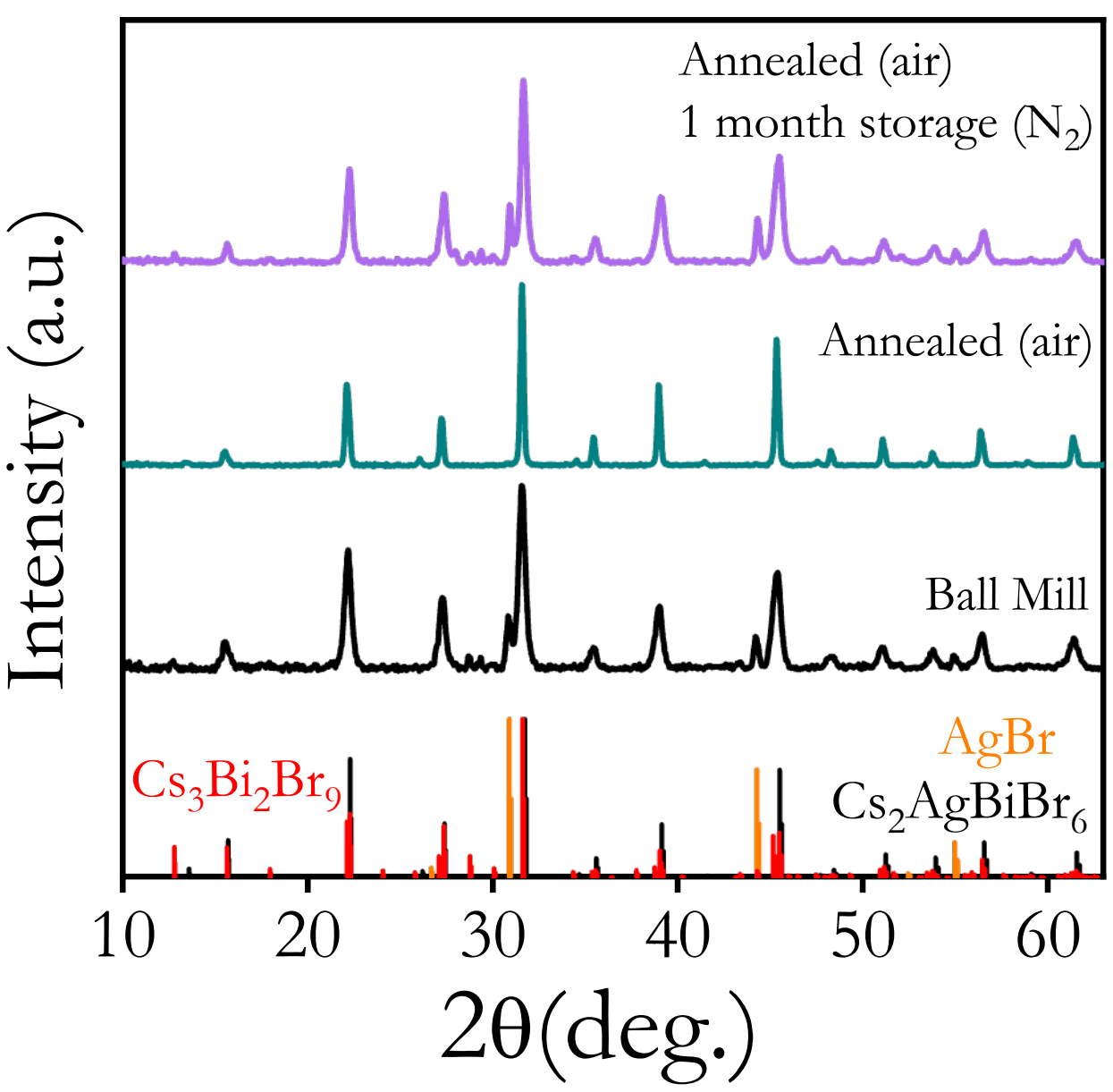

**Figure S13.** X-ray diffraction (XRD) patterns of $Cs_2AgBiBr_6$ powders: as-milled precursor (black), after annealing in air at 250 °C and cooling to room temperature (cyan), and after one month of storage in a glovebox following air annealing (purple). The as-milled powder and the air-annealed sample after glovebox storage show additional reflections attributable to secondary phases, including AgBr and $Cs_3Bi_2Br_9$.

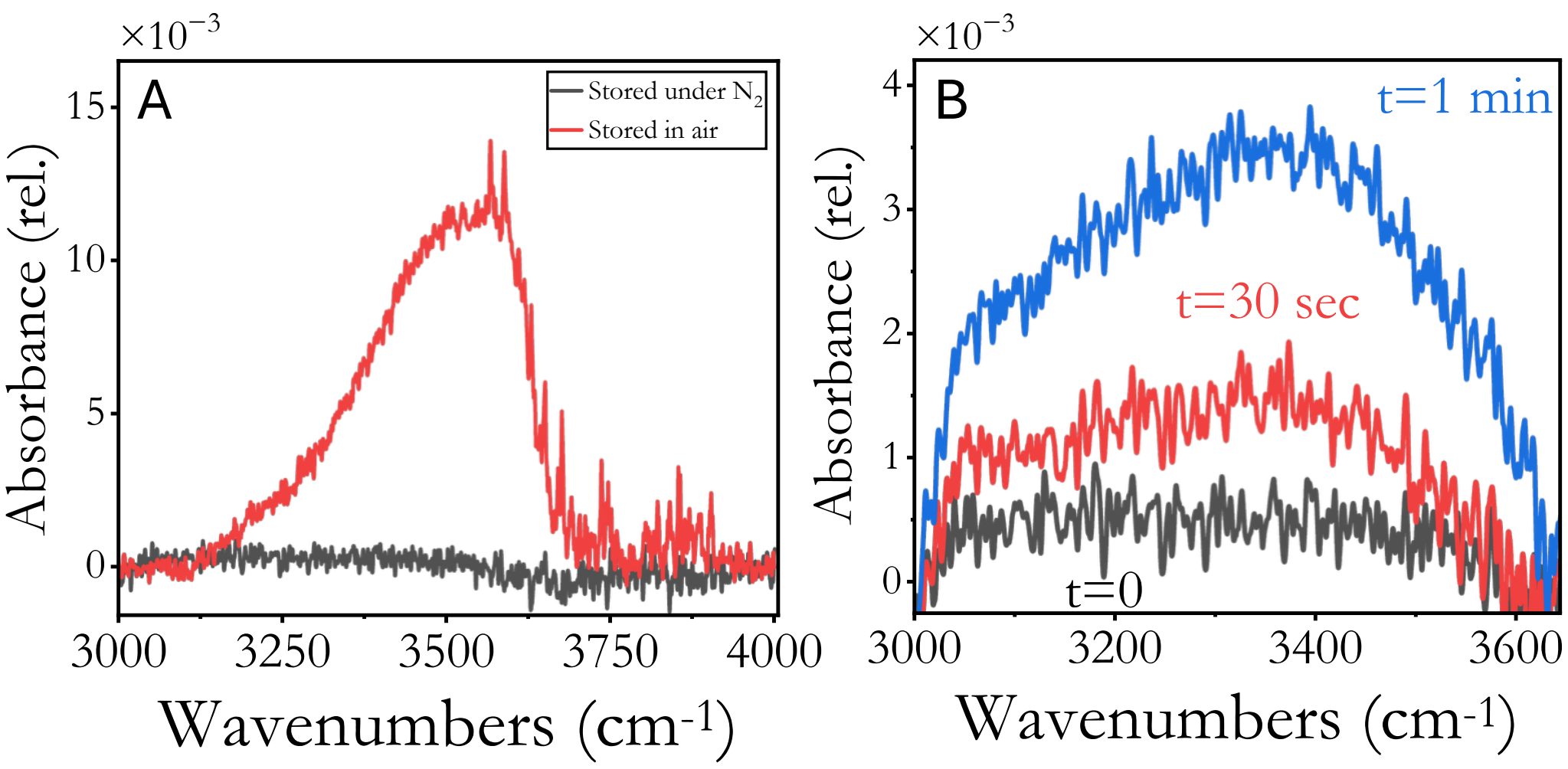

**Figure S14. (A)** FTIR spectra of $Cs_2AgBiBr_6$ in the 3000–4000 $cm^{-1}$ region after different exposure conditions. The black trace is from a sample annealed under $N_2$, cooled to room temperature, and stored in a glovebox. The red trace is from a sample annealed under $N_2$ and subsequently stored in air. The broad absorption band in this region is characteristic of O–H stretching vibrations and is consistent with adsorption of moisture from the ambient air. **(B)** FTIR spectra of a $Cs_2AgBiBr_6$ powder annealed under $N_2$ and stored in a glovebox, measured as a function of time exposed to air (during measurement). For this measurement, a mull was prepared by mixing this powder with a chlorotrifluoroethylene polymer mulling agent (Fluorolube) prior to removing the sample from inert atmosphere for the FTIR and briefly exposed to air during measurement. Spectra collected at t = ~0 s, 30 s, and 1 min in air show rapid growth of the O–H stretching band, indicating rapid moisture adsorption despite the hydrophobic mulling agent.

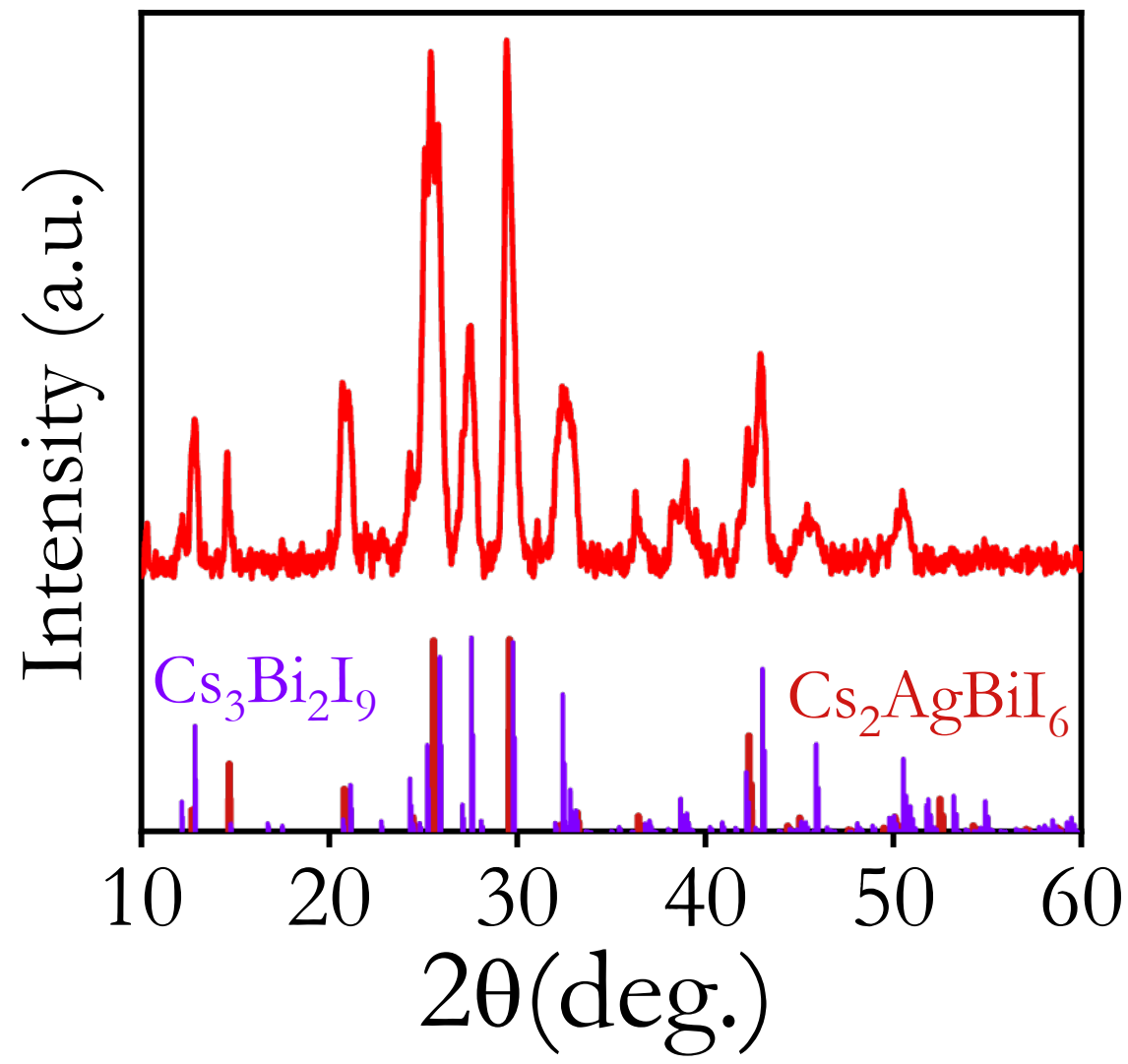

**Figure S15.** X-ray diffraction (XRD) pattern showing the failure to obtain phase-pure $Cs_2AgBiI_6$ thin films following air exposure of the $Cs_2AgBiBr_6$ precursor. The films were initially annealed under inert atmosphere, subsequently exposed to ambient air for one week, and then returned to the glovebox and held overnight under static vacuum before undergoing the anion-exchange reaction with TMSI vapor. The resulting XRD pattern exhibits features consistent with phase segregation, showing reflections attributable to $Cs_3Bi_2I_9$ alongside residual $Cs_2AgBiI_6$, indicating incomplete anion exchange and impurity phase formation.

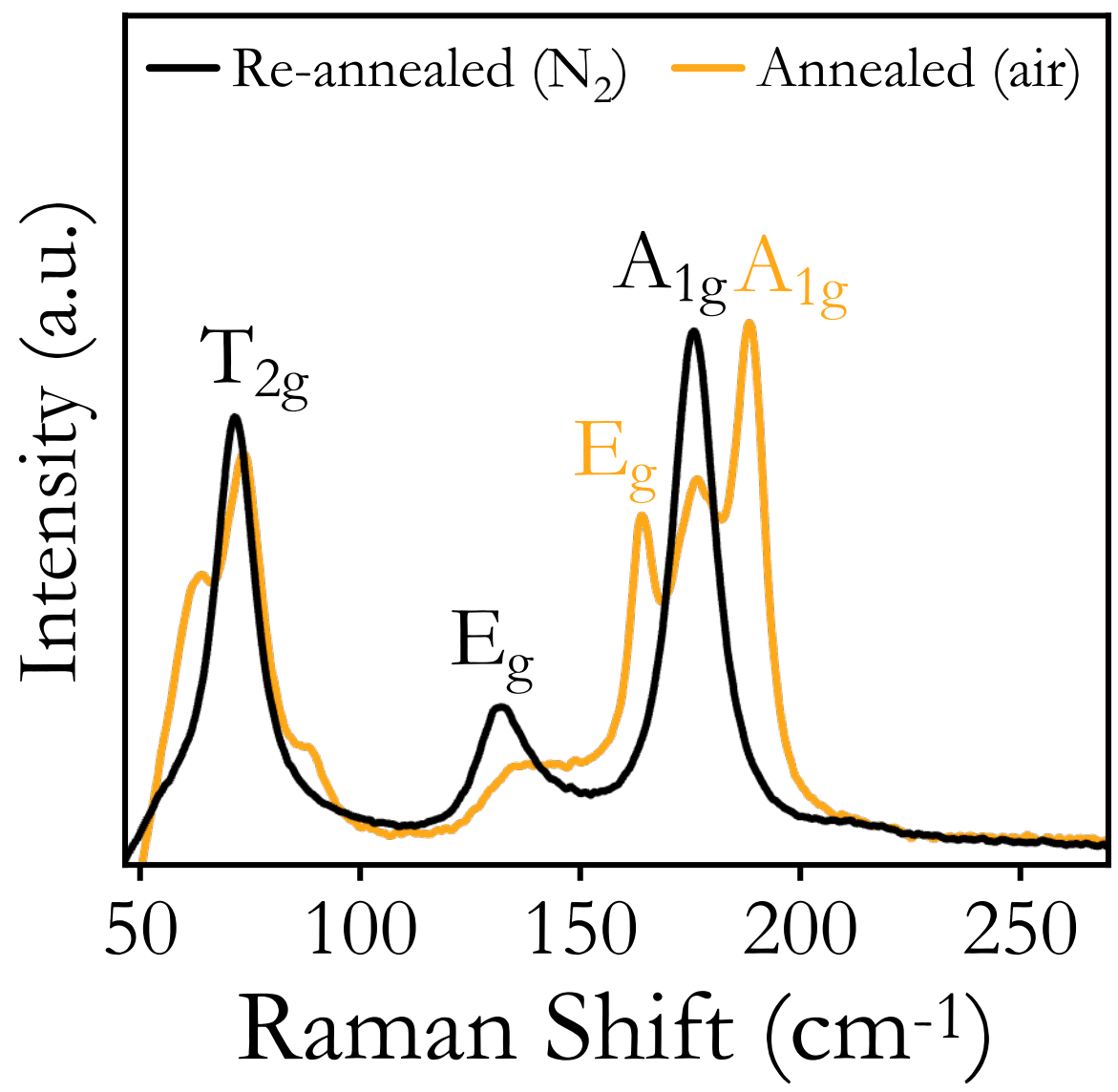

**Figure S16.** Raman spectra of $Cs_2AgBiBr_6$ powders annealed in air and subsequently stored in a glovebox (orange), and after re-annealing the same sample under nitrogen atmosphere (black). The $A_{1g}$, $E_g$, and $T_{2g}$ modes observed after nitrogen re-annealing are characteristic of phase-pure $Cs_2AgBiBr_6$, whereas the additional $A_{1g}$ and $E_g$ features present in the air-annealed sample are assigned to the in-plane and out-of-plane vibrational modes of $[BiBr_6]^{3-}$ octahedra in $Cs_3Bi_2Br_9$.

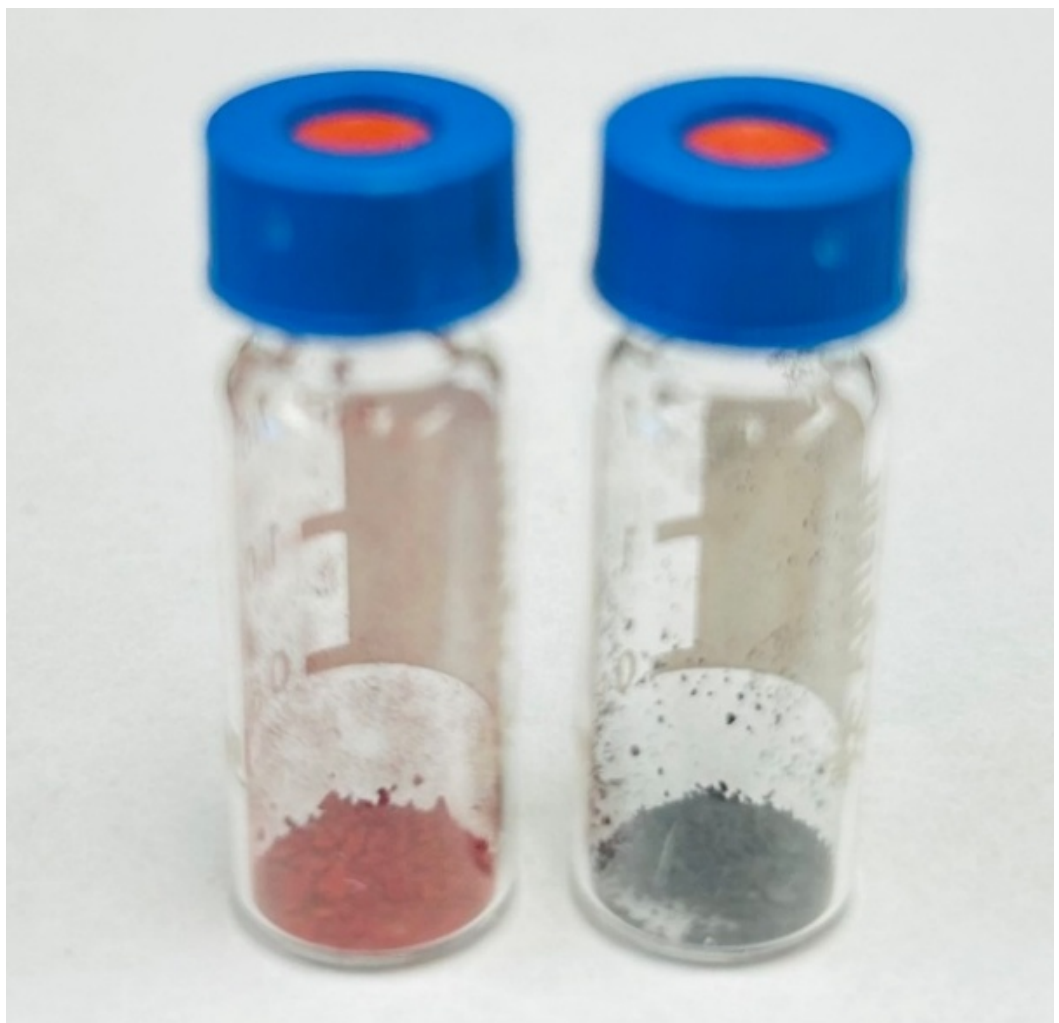

**Figure S17.** (Left) Red powder corresponding to the final product obtained from $Cs_2AgBiBr_6$ annealed in ambient air, transferred into a glovebox, and subsequently exposed to TMSI. The red coloration closely resembles that of $Cs_3Bi_2I_9$. (Right) Black powder corresponding to $Cs_2AgBiBr_6$ annealed in air, transferred into a glovebox, re-annealed under $N_2$ at 250 °C, and then exposed to TMSI. The black coloration reflects the intrinsic color of $Cs_2AgBiI_6$.

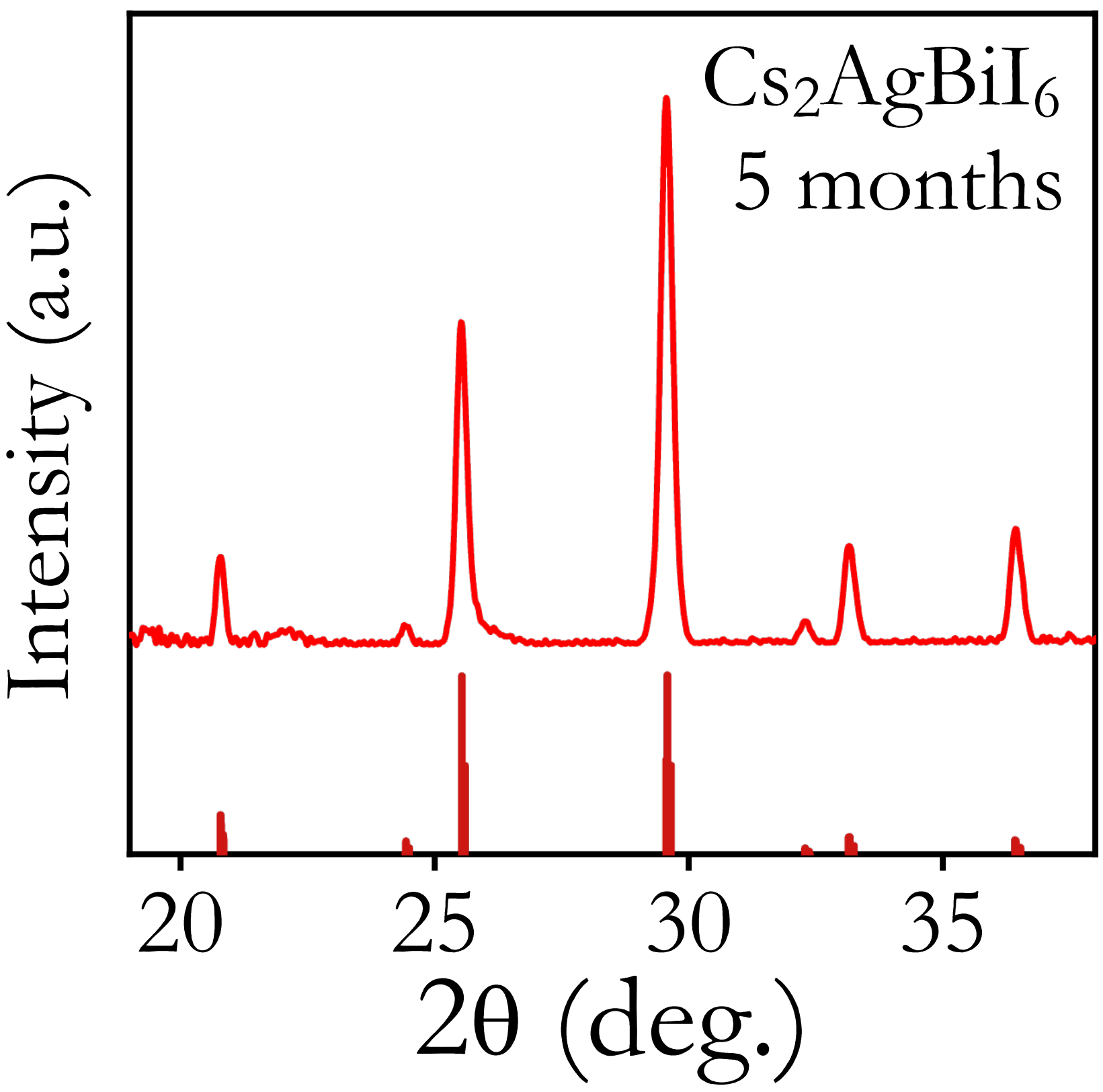

**Figure S18.** XRD pattern of $Cs_2AgBiI_6$ powder after five months of storage under an inert $N_2$ atmosphere, showing no evidence of degradation or impurity phase formation.

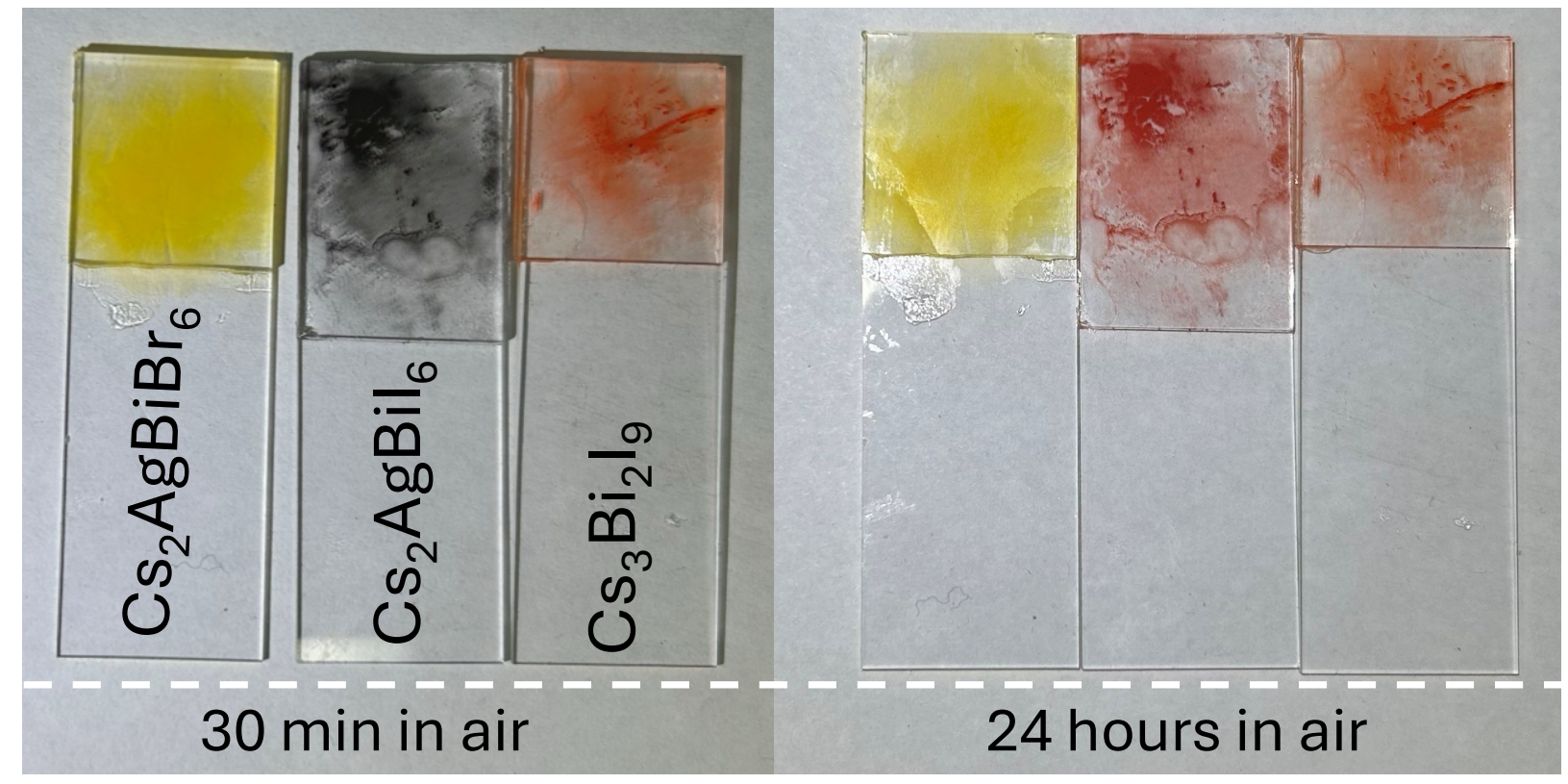

**Figure S19.** Powder samples of $Cs_2AgBiBr_6$, $Cs_2AgBiI_6$, and $Cs_3Bi_2I_9$ prepared for diffuse reflectance measurements. The powders were placed between two glass slides, with fresh samples shown on the left and samples after 24 h of air exposure on the right. Notably, $Cs_2AgBiI_6$ undergoes a color change from black to red upon air exposure, indicating decomposition to the $Cs_3Bi_2I_9$ phase, whereas $Cs_2AgBiBr_6$ and $Cs_3Bi_2I_9$ remain stable.

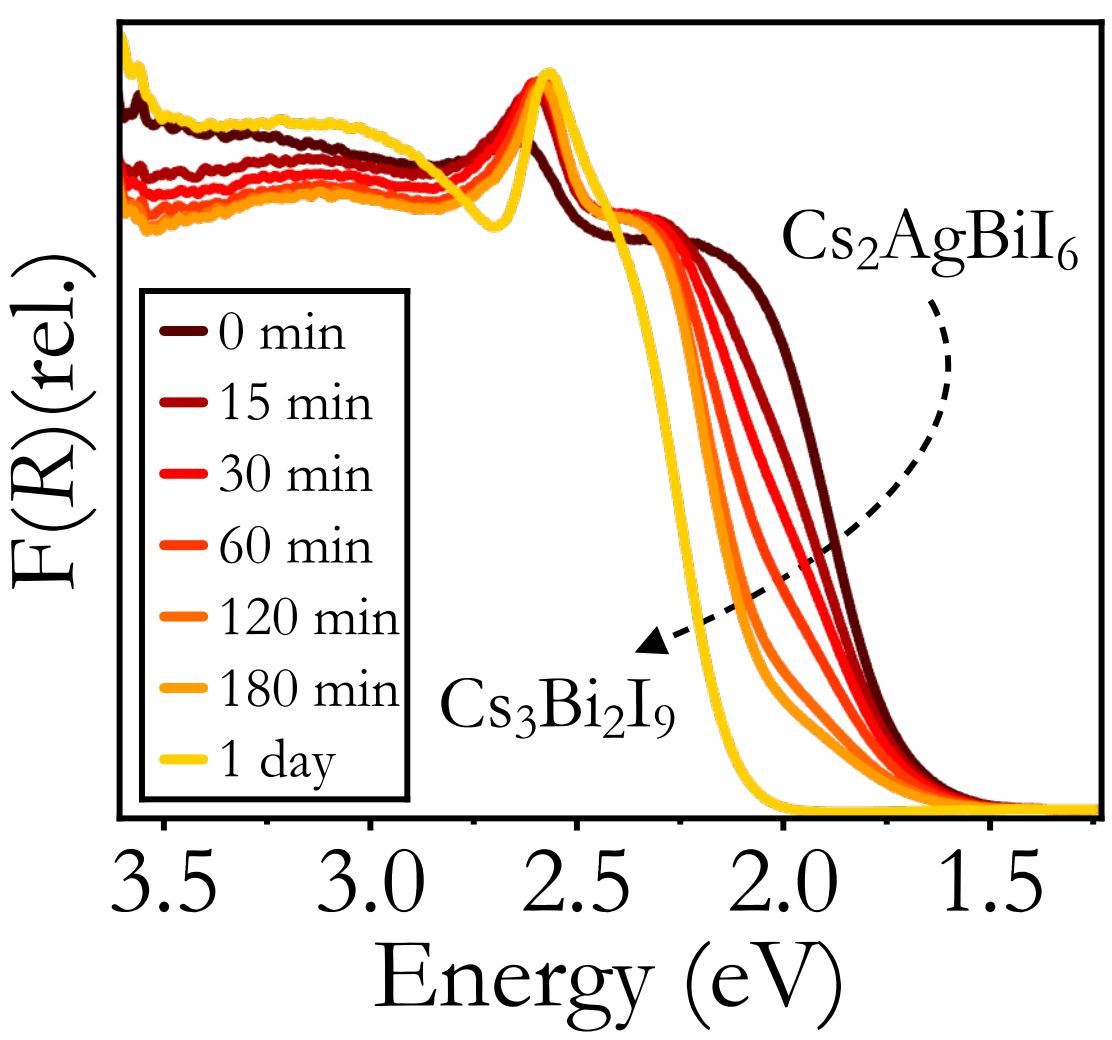

**Figure S20.** Diffuse reflectance F(*R*) spectra of $Cs_2AgBiI_6$ during air exposure, showing a progressive blue shift of the absorption edge consistent with decomposition to the $Cs_3Bi_2I_9$ phase.

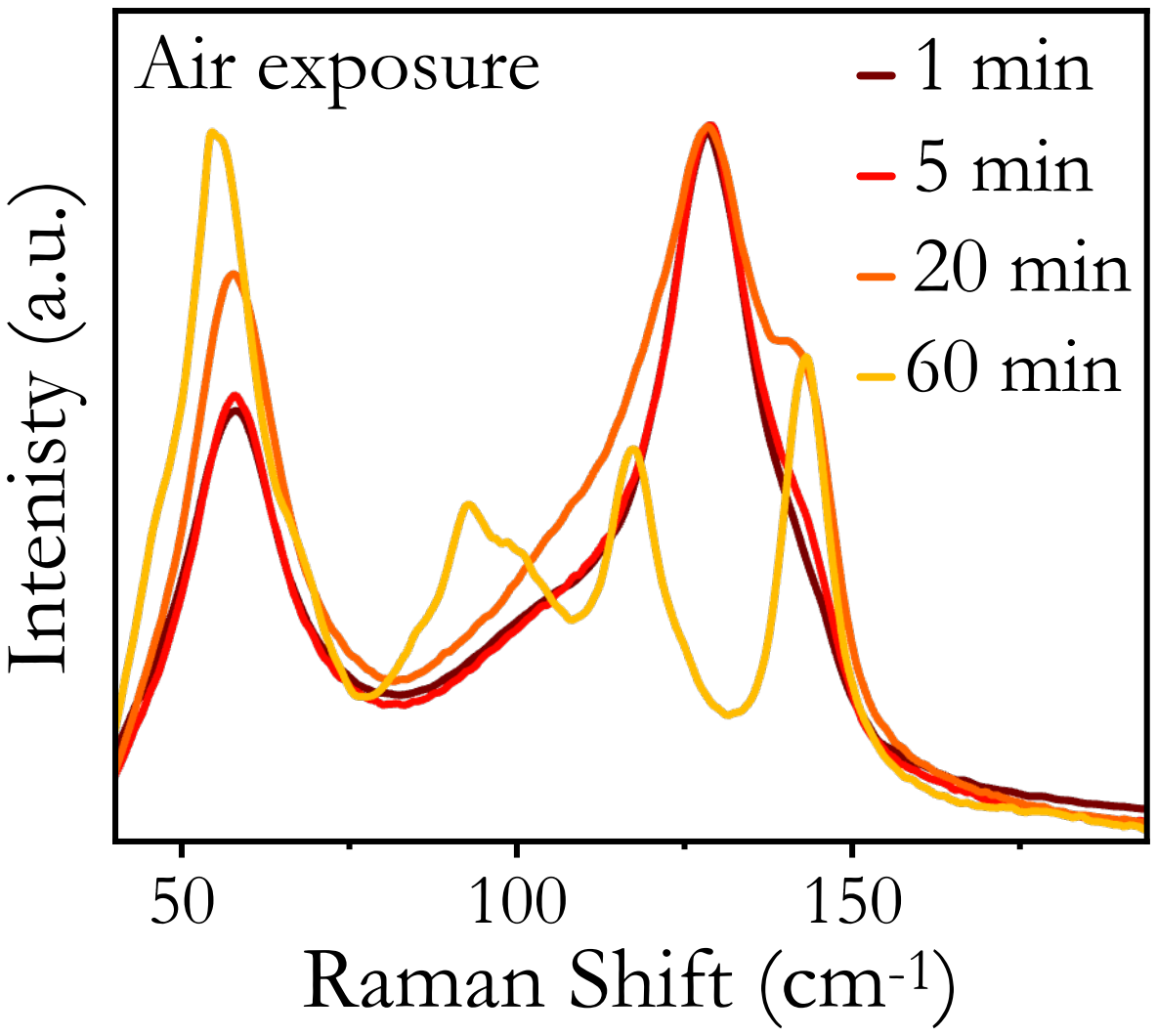

**Figure S21.** Raman spectra of $Cs_2AgBiI_6$ powder collected during air exposure, collected using 785 nm excitation, showing the transformation of $[BiI_6]^{3-}$-related vibrations (deep red) into $Bi_2I_9^{3-}$ vibrations (orange).

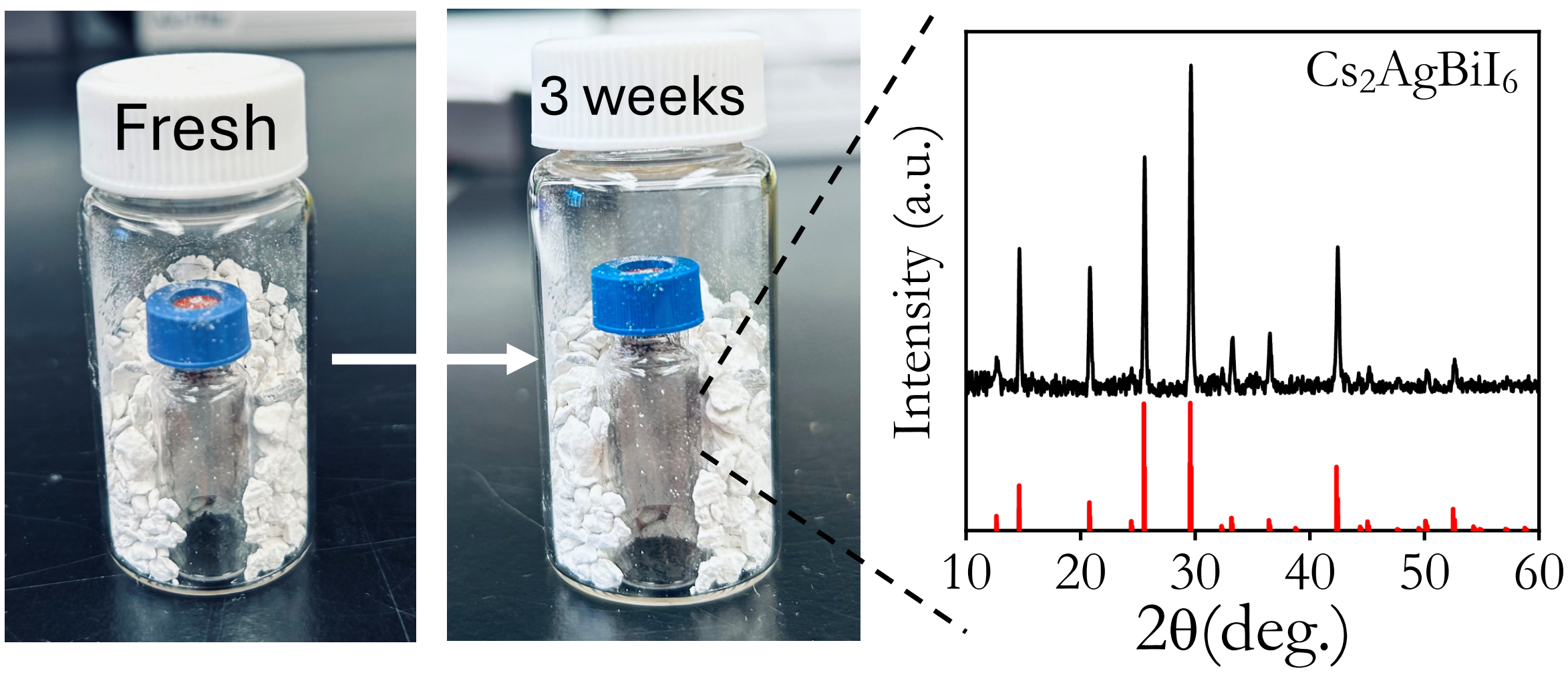

**Figure S22.** Photographs of a microcrystalline $Cs_2AgBiI_6$ sample immediately after preparation (left) and after 3 weeks of storage in air with a desiccant (center). To store in air, the white cap was left off for one hour on the benchtop before sealing, and the blue cap was not sealed but was kept loose to allow mixing of the headspace atmosphere. The latter sample shows no discernible visual signs of degradation or color change. (right) Powder X-ray diffraction (XRD) data for the $Cs_2AgBiI_6$ sample stored in air in the presence of calcium sulfate desiccant for 3 weeks, confirming retention of the phase-pure $Cs_2AgBiI_6$ elpasolite structure.

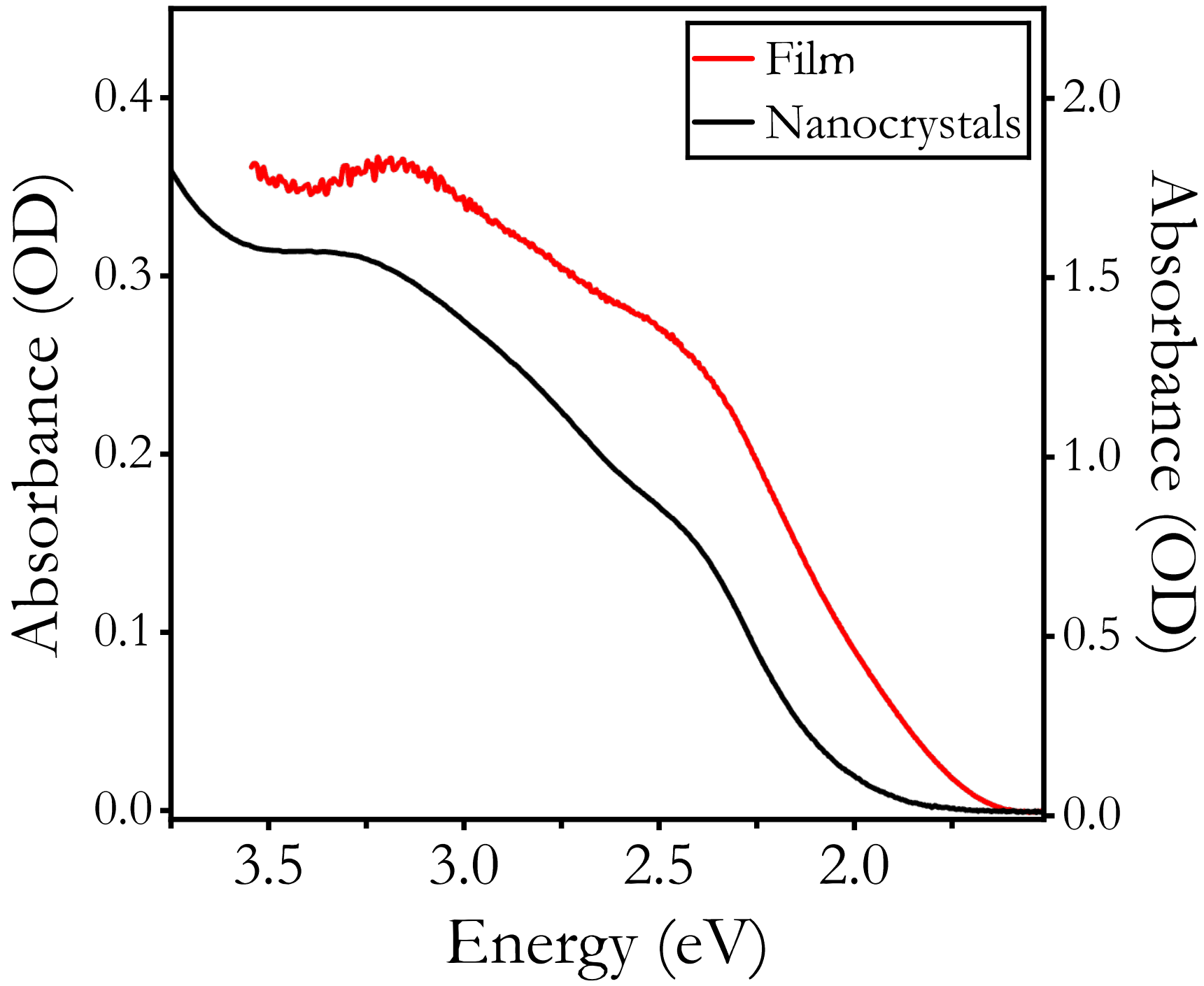

**Figure S23**. Comparison of the room-temperature absorption spectrum of a $Cs_2AgBiI_6$ thin film on a fused silica substrate with that of colloidal $Cs_2AgBiI_6$ nanocrystals in hexane solvent.

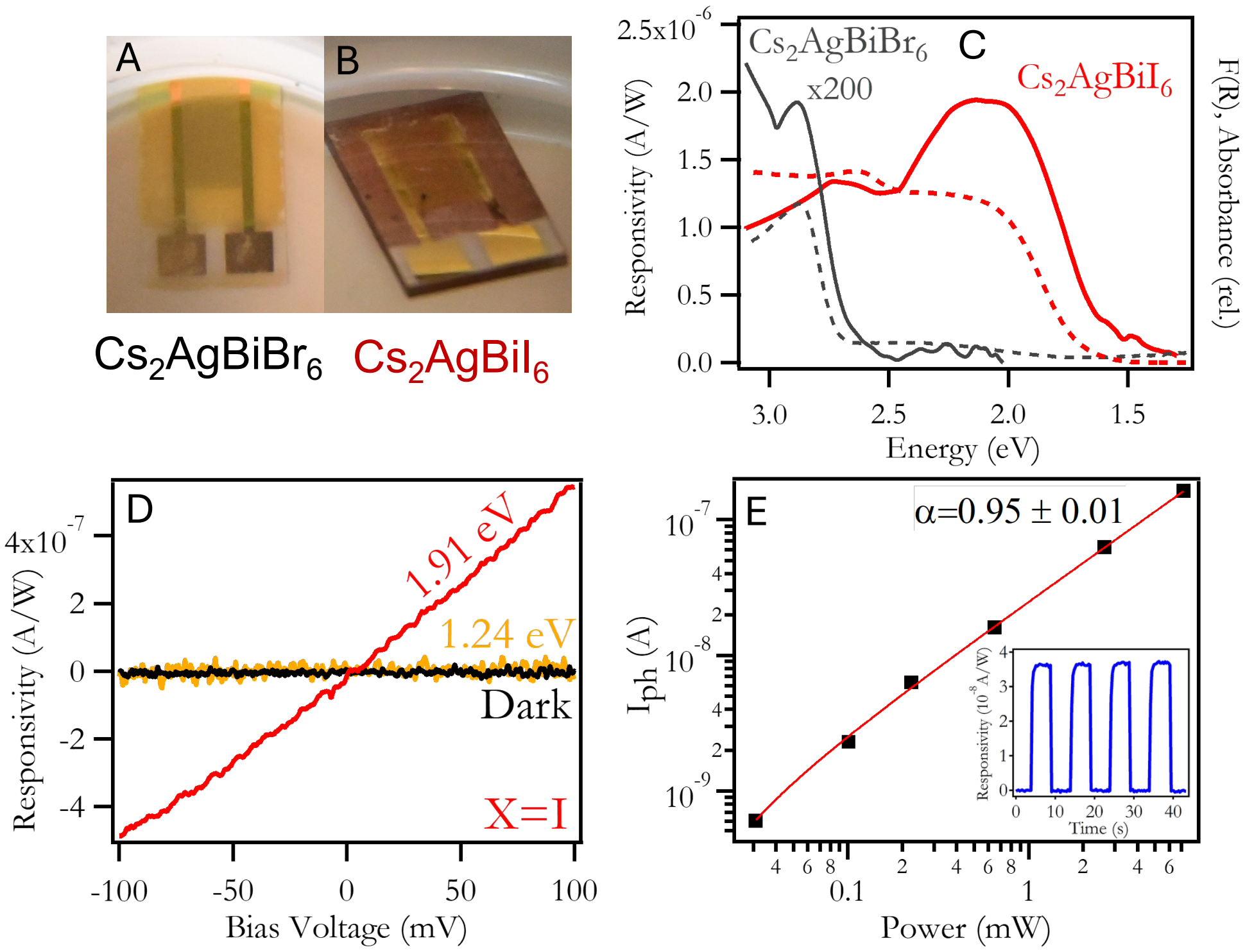

**Figure S24. (A, B)** Photographs of $Cs_2AgBiBr_6$ and $Cs_2AgBiI_6$ films deposited onto fused silica substrates patterned with interdigitated gold electrodes (20 µm spacing). **(C)** Room-temperature photoconductivity action spectra (solid lines) and absorption/pseudo-absorption spectra (dashed lines) of $Cs_2AgBiBr_6$ (black) and $Cs_2AgBiI_6$ (red). These films were prepared simultaneously side-by-side by thermal evaporation, with only one reacted with TMSI, to ensure the same sample thickness. The pseudo-absorption spectrum of the $Cs_2AgBiI_6$ film was obtained *via* transformation of diffuse-reflectance data, and the absorption spectrum of the $Cs_2AgBiBr_6$ film was measured directly. The relative amplitudes of the (pseudo)absorption spectra at their first maxima were estimated from independently measured absorption spectra of nanocrystals at the identical concentration (Figure S28). Photoconductivity measurements for $Cs_2AgBiI_6$ and $Cs_2AgBiBr_6$ films were collected under 50 mV and 2 V bias voltages, respectively. The $Cs_2AgBiBr_6$ photoconductivity is weak and has been multiplied by a factor of 200 for clarity. **(D)** Responsivity–voltage characteristics of the same $Cs_2AgBiI_6$ device under photoexcitation at 1.91 and 1.24 eV (~5 µW cm$^{-2}$), in comparison with the dark curve. All measurements were performed using lock-in detection. **(E)** Photocurrent as a function of incident optical power (532 nm), showing a nearly linear dependence. The red line represents a power-law fit ($I_{ph} \propto P^{\alpha}$) with $\alpha = 0.95$. An uncertainty of ± 0.01 is estimated. Inset: On/off photo-switching of the $Cs_2AgBiI_6$ photoconductivity using 532 nm illumination and a 50 mV bias voltage.

**Comparison between $Cs_2AgBiI_6$ and $Cs_2AgBiBr_6$ photoconductivity using gold electrodes:** Despite a 40× lower applied bias, the $Cs_2AgBiI_6$ film shows 200× *greater* responsivity. To account for the difference in applied bias, we define a voltage-normalized photoconductive response, $\Phi_{\rm ph} = \frac{R}{V}$, justified by the linearity of responsivity–voltage measurements (panel D). Correcting for this difference in bias yields an intrinsic photoconductivity ratio given by $\frac{\Phi_{ph}^{I}}{\Phi_{ph}^{Br}} = \frac{R_{\rm I} V_{\rm Br}}{R_{\rm Br} V_I}$, the value of which is ~4000 for the responses at the first maximum of each compound.

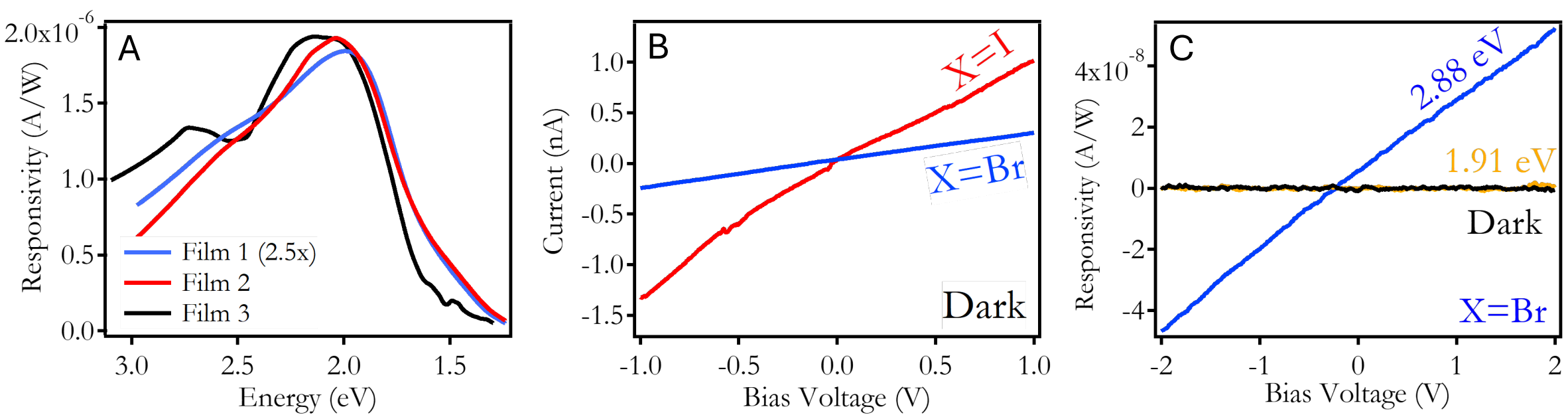

**Figure S25. (A)** Spectral responsivities of three independent $Cs_2AgBiI_6$ thin-film devices with gold electrodes measured at 50 mV bias, demonstrating a reproducible photoconductivity onset at ~1.7 eV. Variations in magnitude arise from differences in chopping frequency and measurement setup, while the spectral features remain consistent across devices. **(B)** Dark current–voltage characteristics of $Cs_2AgBiI_6$ (X = I) and $Cs_2AgBiBr_6$ (X = Br) thin films, both exhibiting linear, symmetric behavior consistent with ohmic contacts; the iodide shows higher dark current. **(C)** Lock-in (photo)responsivity–voltage characteristics of a $Cs_2AgBiBr_6$ device, showing negligible photoresponse below the absorption edge (1.91 eV) and a measurable signal only above it (2.88 eV), confirming that the spectral response reflects photon absorption.

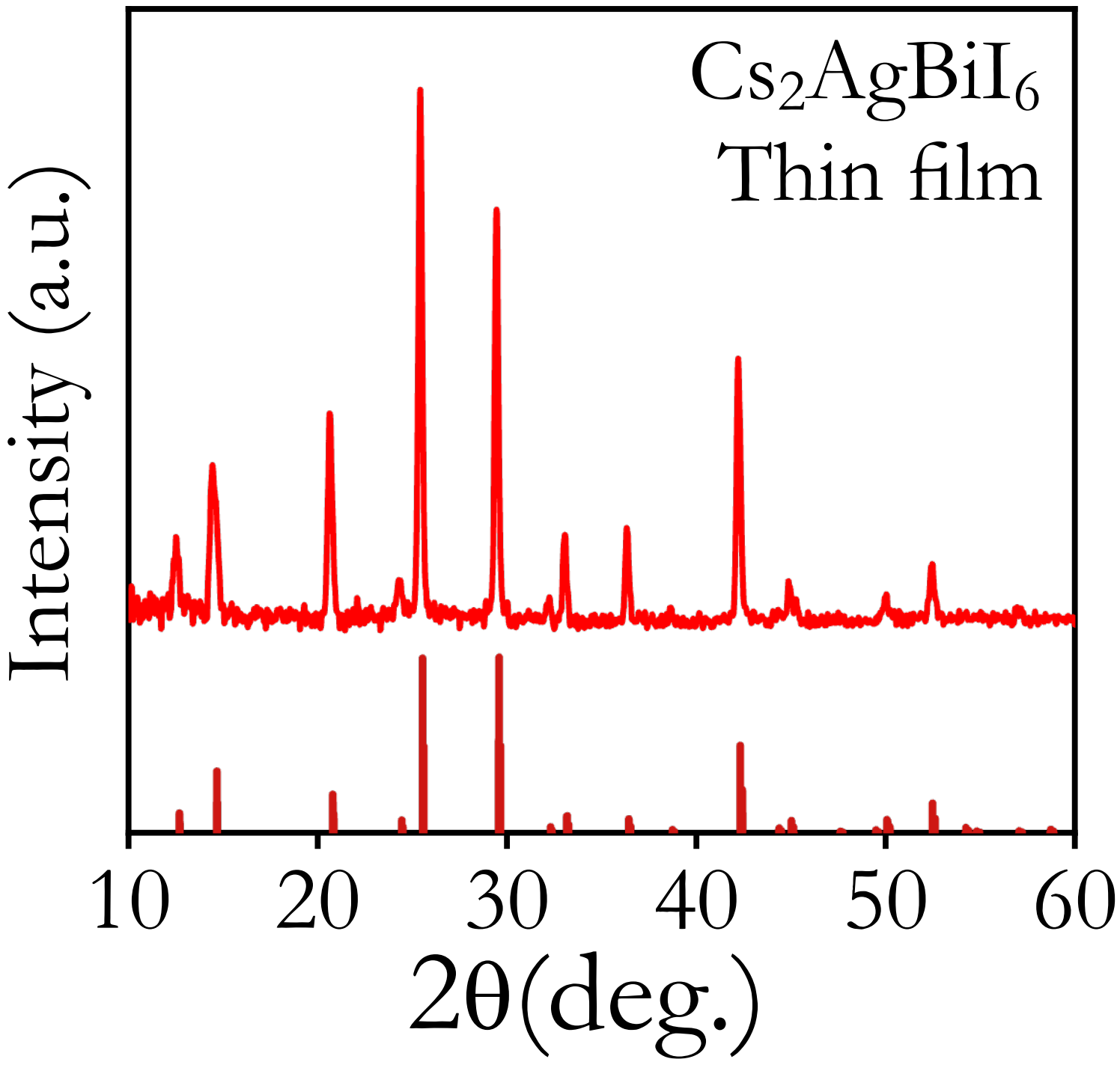

**Figure S26.** XRD pattern of a $Cs_2AgBiI_6$ thin film prepared on a Si substrate at the same time as the device films on gold electrodes.

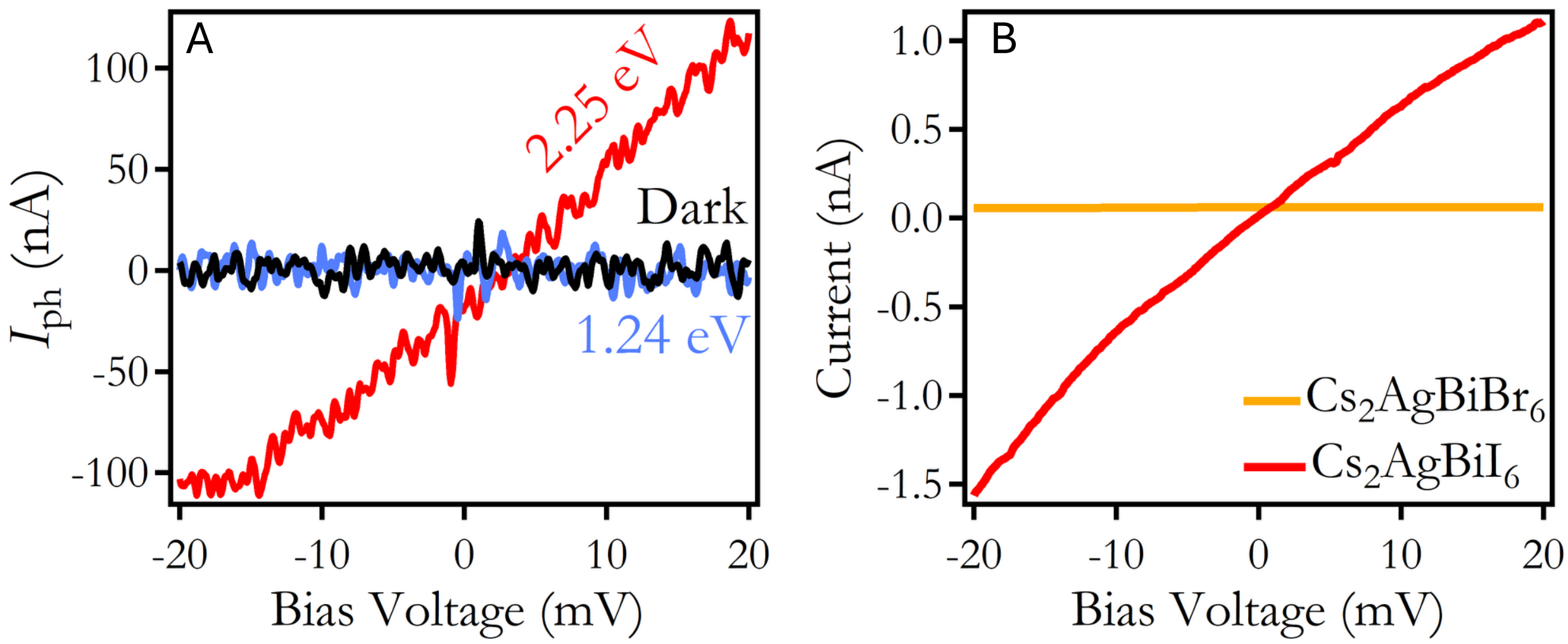

**Figure S27**. **(A)** Photocurrent–voltage characteristics of a representative $Cs_2AgBiI_6$ photoconductivity device with glassy carbon electrodes, measured under photoexcitation at 2.25 eV and 1.24 eV. The dark curve is shown for comparison. All measurements were performed using lock-in detection. **(B)** Dark current–voltage characteristics of $Cs_2AgBiI_6$ and $Cs_2AgBiBr_6$ films, both exhibiting linear, symmetric behavior consistent with ohmic contacts; the iodide shows higher dark current.

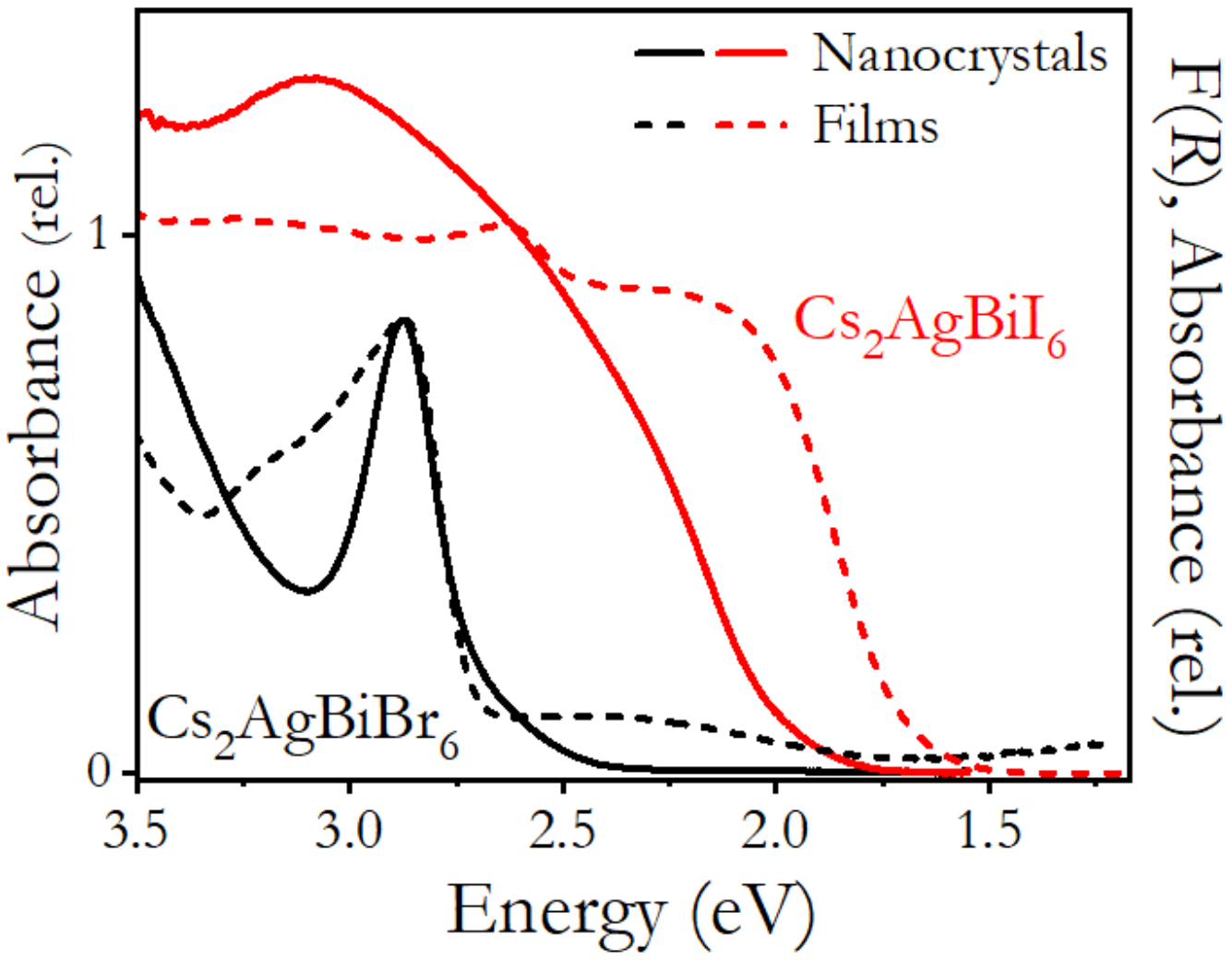

**Figure S28.** Absorption and pseudo-absorption spectra of $Cs_2AgBiBr_6$ (black dashed) and $Cs_2AgBiI_6$ (red dashed) thin films, compared with absorption spectra of colloidal $Cs_2AgBiBr_6$ (black solid) and $Cs_2AgBiI_6$ (red solid) nanocrystals collected at an identical nanocrystal concentration. The $Cs_2AgBiBr_6$ film spectrum is plotted as absorbance, and the $Cs_2AgBiI_6$ film spectrum is presented as the Kubelka–Munk–transformed diffuse reflectance, F(R) (pseudo-

absorption). The relative amplitudes of the $Cs_2AgBiBr_6$ and $Cs_2AgBiI_6$ film spectra at their absorption edges were scaled to roughly match the relative amplitudes of the corresponding nanocrystal absorption spectra at their absorption edges.

**Table S1. EDX quantitative analysis of the compositions of $Cs_2AgBiX_6$ (X = Br, I) powders and evaporated thin films, averaged over five distinct regions per sample. Blank cells mean no detectable signal for this element.**

| | Powder | | Film | |
|---|---|---|---|---|
| Elements | X = Br | X = I | X = Br | X = I |
| Cs | 0.19 | 0.22 | 0.16 | 0.21 |
| Ag | 0.10 | 0.09 | 0.11 | 0.08 |
| Bi | 0.07 | 0.07 | 0.09 | 0.10 |
| Br | 0.64 | | 0.64 | |
| I | | 0.62 | | 0.61 |

**Table S2. Fitting results from PL decay curves measured for bulk $Cs_2AgBiI_6$ powder.**

| Temperature (K) | $A_1$ | $\tau_1$ (μs) | $A_2$ | $\tau_2$ (μs) | $A_3$ | $\tau_3$ (μs) | $\tau_{avg}$ (μs) |
|---|---|---|---|---|---|---|---|
| 4 | 0.73 | 2.3 | 0.24 | 14.7 | 0.04 | 126.0 | 10.7 |
| 10 | 0.74 | 2.4 | 0.24 | 14.6 | 0.05 | 124.6 | 10.7 |
| 20 | 0.74 | 2.5 | 0.23 | 15.7 | 0.05 | 130.1 | 11.2 |
| 30 | 0.74 | 2.5 | 0.24 | 15.8 | 0.05 | 132.6 | 11.6 |
| 40 | 0.72 | 2.5 | 0.25 | 15.4 | 0.05 | 132.8 | 12.0 |
| 50 | 0.74 | 2.5 | 0.24 | 16.5 | 0.05 | 141.7 | 12.3 |
| 60 | 0.74 | 2.6 | 0.24 | 17.1 | 0.05 | 148.3 | 13.0 |
| 70 | 0.73 | 2.5 | 0.24 | 16.7 | 0.05 | 147.7 | 13.2 |
| 80 | 0.73 | 2.4 | 0.24 | 16.5 | 0.05 | 149.4 | 13.2 |
| 100 | 0.72 | 2.1 | 0.24 | 15.3 | 0.05 | 145.5 | 12.9 |
| 120 | 0.71 | 1.7 | 0.24 | 12.9 | 0.05 | 133.0 | 11.4 |
| 140 | 0.72 | 1.3 | 0.23 | 10.2 | 0.05 | 98.9 | 8.2 |
| 160 | 0.65 | 0.5 | 0.29 | 3.5 | 0.05 | 37.3 | 3.3 |
| 180 | 0.66 | 0.4 | 0.29 | 2.8 | 0.05 | 27.6 | 2.4 |
| 200 | 0.66 | 0.4 | 0.28 | 2.1 | 0.05 | 17.0 | 1.7 |
| 220 | 0.61 | 0.9 | 0.33 | 0.0 | 0.03 | 17.9 | 1.2 |
| 250 | 0.83 | 0.3 | 0.15 | 2.0 | | | 0.6 |
| 270 | 0.88 | 0.3 | 0.12 | 1.6 | | | 0.4 |

$\tau_{avg}$ is calculated from the tri-exponential fitting parameters using the weighted-average equation:

$\tau_{avg} = \frac{\Sigma A_i \tau_i}{\Sigma A_i}$.